\documentclass[10pt,superscriptaddress,aps,prl,twocolumn,showpacs,floatfix, nobalancelastpage]{revtex4-2}
\usepackage[utf8]{inputenc}
\usepackage{graphicx,amsmath,amsfonts,amssymb}
\usepackage[english]{babel}
\usepackage{color}	

\usepackage{amsmath}
\usepackage{physics}
\usepackage{graphicx} 
\usepackage[colorlinks=true, allcolors=blue]{hyperref}
\usepackage{braket}
\usepackage{tcolorbox} 
\usepackage{dsfont}
\usepackage{bm}

\newcommand{\llangle}{\langle\!\langle}
\newcommand{\rrangle}{\rangle\!\rangle }
\renewcommand{\H}{\mathcal{H}}
\renewcommand{\L}{\bm{\mathcal{L}}}
\newcommand{\D}{\mathcal{D}}
\newcommand{\F}{\mathcal{F}}

\newcommand{\orcid}[1]{\href{https://orcid.org/#1}{\includegraphics[width=7pt]{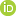}}}

\begin{document}
\preprint{APS/123-QED}

\title{Experimental observation and application of the genuine Quantum Mpemba Effect}

\author{\mbox{Bruno P. Schnepper}\orcid{0000-0003-1159-3166}}
\affiliation{Centro de Ci\^{e}ncias Naturais e Humanas, Universidade Federal do ABC, Avenida dos Estados 5001, 09210-580 Santo Andr\'e, S\~{a}o Paulo, Brazil}

\author{\mbox{Jefferson L. D. de Oliveira}\orcid{0000-0003-3743-4841}}
\affiliation{Centro de Ci\^{e}ncias Naturais e Humanas, Universidade Federal do ABC, Avenida dos Estados 5001, 09210-580 Santo Andr\'e, S\~{a}o Paulo, Brazil}

\author{\mbox{Yan A. C. Avó}\orcid{0000-0002-2474-3947}}
\affiliation{Centro de Ci\^{e}ncias Naturais e Humanas, Universidade Federal do ABC, Avenida dos Estados 5001, 09210-580 Santo Andr\'e, S\~{a}o Paulo, Brazil}

\author{\mbox{Carlos H. S. Vieira\orcid{0000-0001-7809-6215}}}
\affiliation{Centro de Ci\^{e}ncias Naturais e Humanas, Universidade Federal do ABC, Avenida dos Estados 5001, 09210-580 Santo Andr\'e, S\~{a}o Paulo, Brazil}
\affiliation{Department of Physics, State Key Laboratory of Quantum Functional Materials, and Guangdong Basic Research Center of Excellence for Quantum Science, Southern University of Science and Technology, Shenzhen 518055, China}

\author{Krissia Zawadzki\orcid{0000-0002-6133-0850}}
\affiliation{Instituto de F\'isica de S\~ao Carlos, Universidade de S\~ao Paulo,
CP 369, 13560-970 S\~ao Carlos, S\~ao Paulo, Brazil}

\author{Roberto M. Serra\orcid{0000-0001-9490-3697}}
\affiliation{Centro de Ci\^{e}ncias Naturais e Humanas, Universidade Federal do ABC, Avenida dos Estados 5001, 09210-580 Santo Andr\'e, S\~{a}o Paulo, Brazil}

\begin{abstract}
Coherence is an inherently quantum property that deeply affects microscopic processes, including thermalization phenomena. A striking example is the quantum Mpemba effect (QME), in which a system can exhibit anomalous relaxation, thermalizing faster from a state initially farther from equilibrium than from one closer. Here, we experimentally investigate the genuine QME and observe how the dynamics of a spin-1/2 system interacting with a heat sink can be sped-up to equilibrium. The mechanism underlying this speed-up is uncovered through detailed analysis of the heat exchange process. Furthermore, we apply the QME in a quantum Otto refrigerator, thereby increasing its cooling power. This proof-of-concept experiment unveils new practical paths for improving quantum thermal tasks. 
\end{abstract}

\maketitle

Accelerating the relaxation of quantum systems towards a steady state can be helpful in certain tasks in non-equilibrium quantum thermodynamics~\cite{Seifert_2012,Goold_2016,Myers2022,Cappellaro2022,Sone2025}, it is also crucial for quantum technologies relying on rapid state preparation~\cite{An_2016,Odelin_2019,TORRONTEGUI_2013}, qubit engineering~\cite{Pan_2022,Bruzewicz_2019,Ladd_2010}, and increasing power in quantum thermal cycles~\cite{Abah_2020,Peterson_2019,Camati_2019,Arısoy_2021}. One possibility to achieve this speed-up is by exploiting the so called quantum Mpemba effect (QME) \cite{Zhiyue2017,Carollo2021,Kumar2024,Liu2024,Ares_2025,Teza2025, Strachan2025}, notorious for its classical counterpart~\cite{aristotle1987,Bacon1620,Descartes1637,Black1775,Mpemba_1969}. This counter-intuitive effect, named in the classical context in 1969, has much older records from many well-known figures in the history of science, including Aristotle, Francis Bacon, and René Descartes~\cite{aristotle1987,Bacon1620,Descartes1637,Black1775,Mpemba_1969}. This phenomenon,  where a “hot” state
tends to cool down faster than its “warm” state, has been recently re-interpreted as an anomalous relaxation occurring whenever a fixed point is reached faster when starting from states further from it. Although related classical experiments date from the late 18th century~\cite{Black1775}, such an effect remains a subject of active research in the classical context, including recent experiments on colloidal systems~\cite{Kumar2020,Kumar2022,Malhotra2024} and the development of theoretical frameworks within non-equilibrium statistical mechanics~\cite{Zhiyue2017,Klich2019,Gal2020,Teza2023}.

In the quantum context, a general theoretical description of this effect in the Markovian
regime was proposed only very recently by Moroder et al.~\cite{Mpemba2024}, from the perspective of non-equilibrium quantum thermodynamics~\cite{Adesso2018,Goold_2016,Anders2016}. In this framework, the authors employ Markovian open dynamics~\cite{Zhiyue2017,Carollo2021,Klich2019}, specifically focusing on relaxation processes described by Davies maps~\cite{Davies1979,Dann2021}. The key feature of Davies maps is that their mathematical structure allows the QME to be described directly from the dynamical generator. Hence, an exponential speed-up towards equilibrium can be achieved by a suitable transformation of the system's initial state, ensuring minimal overlap with the slowest-decaying modes of the dissipative dynamics. It has been shown that the transformed initial state possesses higher non-equilibrium free energy ~\cite{Donald1987,Parrondo2015}, which has been argued to be a suitable quantity for measuring the distance from the fixed point, regardless of whether the initial state is thermal or not. In this scenario, following the relaxation dynamics through the non-equilibrium free energy, one observes that the transformed state reaches equilibrium faster than the original state (closer to equilibrium).   
 
Nonetheless, experimental implementations of the QME present some challenges, mostly stemming from the need for high-level quantum control and complete characterization of the relaxation dynamics. Initial attempts to experimentally investigate some versions of the QME have been made recently, employing single trapped ions~\cite{sME_exp2025}. Here, we use a spin-1/2 system and Nuclear Magnetic Resonance (NMR) techniques~\cite{Levitt2008,Sarthour2007, Jones2011, Vieira2023,Dawei2022} to advance the experimental investigation of the genuine QME. The key mechanism of the genuine QME is also uncovered through the measurement of the contribution to the dynamics of each Lindbladian eigenmode during the heat exchange, which is presented in the End Matter. We also exploit the genuine QME in a quantum refrigeration cycle~\cite{He2002,Kosloff2014,Maslennikov2019} to increase its cooling power. It provides a proof-of-concept experiment demonstrating the possibility of harnessing the QME to gain an advantage in practical applications.

\textit{Theoretical framework}\textemdash The QME emerges from the properties of a system undergoing Markovian dissipative dynamics. In Liouville space, this evolution can be described by the master equation~\cite{Fano1964,Gyamfi_2020}
\begin{equation}\label{eq::lindblad}
    \frac{d}{dt}|\hat\rho(t)\rrangle = \L|\hat\rho(t)\rrangle,
\end{equation}
where $|\hat\rho(t)\rrangle$ is the vectorized density operator and $\L$ is the Lindbladian super-operator (the dynamics generator). Considering a general time-independent Hamiltonian of dimension $\dim(\H) = d$, a formal solution of Eq.~\eqref{eq::lindblad} can be expressed in terms of the eigen-decomposition of $\L$ (see Supplemental Material)
\begin{equation}\label{eq::lindblad_sol}
  |\hat\rho(t)\rrangle = \sum_{k=1}^{d^2}e^{t\lambda_k}|\hat\zeta_k\rrangle\llangle \hat\xi_k|\hat\rho(0)\rrangle,
\end{equation}
with eigenvalues $0 = \lambda_1 \leq |\Re (\lambda_2)| \leq ... \leq |\Re (\lambda_{d^2})|$, left and right eigenvectors $\{\llangle \hat\xi_k|\}$ and $\{|\hat\zeta_k\rrangle\}$ satisfying $\llangle \hat\xi_k|\L = \lambda_k\llangle \hat\xi_k|$ and $\L|\hat\zeta_k\rrangle = \lambda_k|\hat\zeta_k\rrangle$, respectively.  Equation~\eqref{eq::lindblad_sol} includes different timescales associated with the complex eigenvalues, $\lambda_k$. In this formalism, the null eigenvalue, $\lambda_1$, is associated with the fixed point, $|\hat \zeta_1\rrangle$, related to the density operator steady state.     The longest timescale in the system's evolution is associated with the \textit{slowest eigenmode}, $\llangle \hat\xi_2|$, such that the norm-1 difference between the evolved state $|\hat\rho(t)\rrangle$ and the steady state $|\hat\zeta_1\rrangle$ is  $||\, |\hat\rho(t)\rrangle - |\hat\zeta_1\rrangle \,||_1 \propto \exp{[\Re(\lambda_2)t]}$. Hence, a speed-up towards the steady state can be achieved by a suitable unitary transformation $\hat U\mapsto \hat\rho_{\text{mb}}(0) \equiv \hat U\hat\rho(0)\hat U^\dagger$ on the initial state, optimally one that removes the overlap between the initial state and the slowest eigenmode of the Lindbladian super-operator, so that
$\llangle \hat\xi_2|\hat\rho_{\text{mb}}(0)\rrangle \approx 0$.

As evidenced in Ref.~\cite{Mpemba2024}, for a particular class of thermalization processes described by Davies maps, the complete Lindbladian represented in the Hamiltonian eigenbasis can be split into population ($p$) and coherence ($c$) subspaces $\L = \L_p \oplus \L_c$. In this way, it can be observed that diagonalizing the system's initial state in the energy eigenbasis eliminates all dynamical terms due to $\L_c$, namely  \(\llangle \hat\xi_j|\hat\rho_{\text{mb}}(0) \rrangle\). Furthermore, it is essential to emphasize that a speedup by itself does not constitute a \textit{genuine} quantum Mpemba effect~\cite{Mpemba2024}. The main feature of the classical Mpemba effect is its tendency to enable faster cooling in systems with higher initial temperatures than in those with lower initial temperatures. In the context of out-of-equilibrium quantum systems, where conventional notions of temperature cannot be ascribed, this phenomenon can be addressed through the non-equilibrium free energy \(\F_{\text{neq}}=\Tr (\hat{H}\hat{\rho}) + 1/\beta \Tr (\hat{\rho}\ln\hat{\rho})\), where \(\beta\) is the inverse temperature of the fixed point. Following this rational, a genuine quantum Mpemba effect~\cite{Mpemba2024} requires that the transformed initial state possesses a higher non-equilibrium free energy $\F_{\text{neq}}(\hat\rho_{\text{mb}}(0)) > \F_{\text{neq}}(\hat{\rho}(0))$ along with an intersection in the thermalization curves, signifying the existence of a time \(\tau'\) (denoting the crossing time) such that \(\F_{\text{neq}}(\hat\rho_{\text{mb}}(t)) < \F_{\text{neq}}(\hat{\rho}(t))\) holds for all times \(t>\tau'\). 

\begin{figure}[t!]
    \centering
    \includegraphics[width=\linewidth]{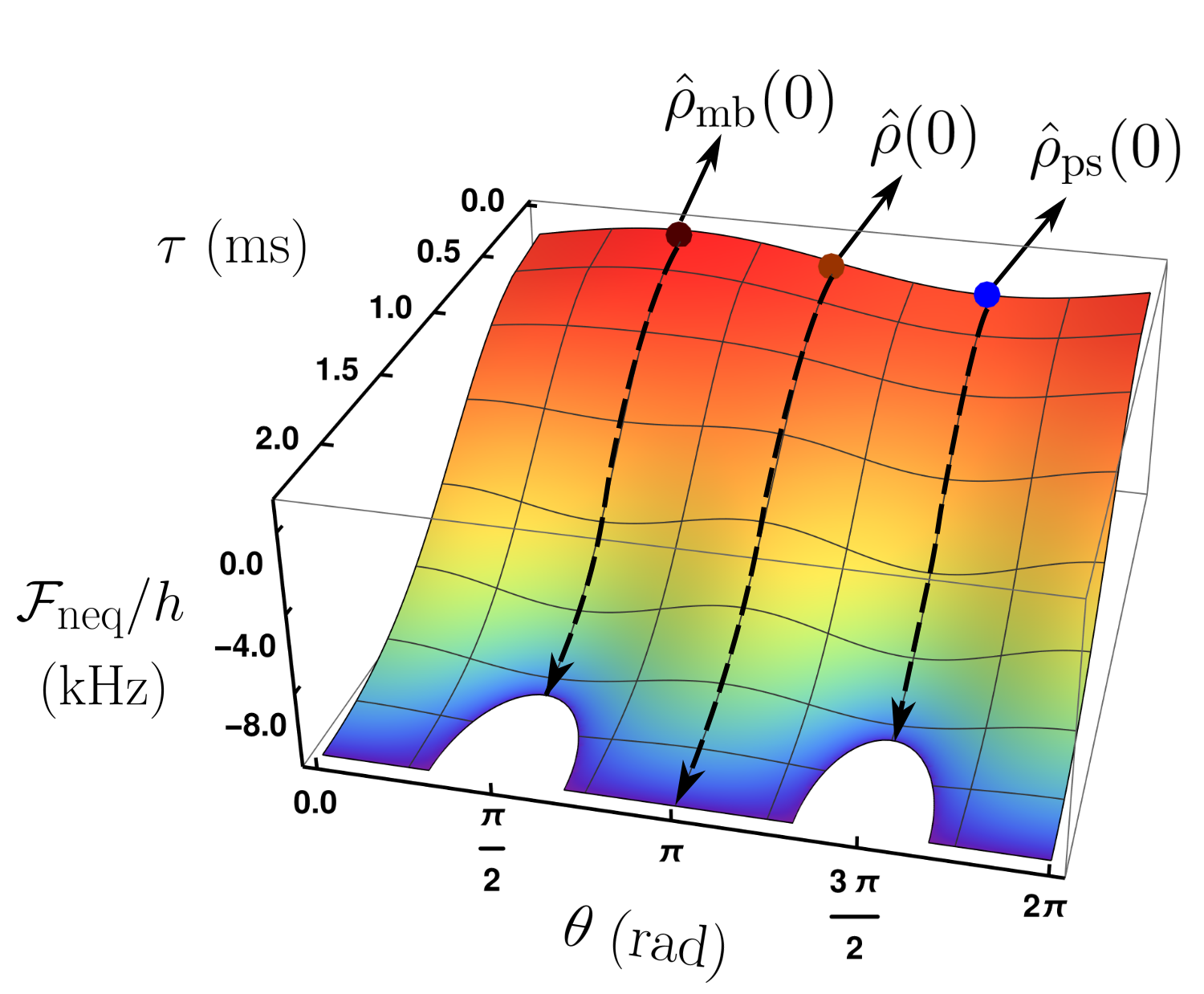}
    \caption{Schematics for the quantum Mpemba effect. Theoretical simulation of the non-equilibrium free energy dynamics, $\F_{\text{neq}}$, in the employed cooling protocol (as described in Fig.~\ref{fig:fig2}). The equilibrium state of this dynamics is the Gibbs state for the Hamiltonian $\hat \H^{\text{eq}} = -2\pi\hbar\nu_1\hat\sigma_z$ at effective temperature $k_{B} T/ h = 6.95 \; \text{kHz}$, where the energy gap is determined by the frequency $\nu_1 = 2\; \text{kHz}$. The blue line represents the system very close to equilibrium such that $\F_{\text{neq}}(\hat\rho_\theta(\tau))-\F_{\text{eq}}<10^{-4}$, it is also the intersection with the lower $x$-$y$ plane defined by $\F_{\text{eq}}/h \approx -11.57$~kHz. See also Supplementary Figs.~S2 and~S3. The following set of initial states is considered, $\hat\rho_\theta(0) = \hat R_y(\theta)\hat\rho (0)\hat R_y(-\theta)$, where $\hat{\rho} (0) = \dyad{x_{+}}$, $|x_{\pm}\rangle$ are the $\hat\sigma_x$ Pauli operator eigenstates (with eigenvalues $\pm 1$) and $ \hat R_y(\theta)$ is the usual SU2 rotation. The angle $\theta$, in the longitudinal axis, is directly related to the projection of the initial state onto the slowest eigenmode of the Lindbladian super-operator, $\llangle\hat\xi_2|\hat\rho_{\theta}(0)\rrangle$, whereas the time duration of the cooling protocol, $\tau$, is represented in the horizontal axis. The initial states indicated by $\hat\rho_{\text{mb}}(0)$ and $\hat\rho(0)$ correspond to those in the experimental realization, and $\hat\rho_{\text{ps}}(0)$ is the passive state of the mentioned Hamiltonian with populations defined by the $\hat{\rho}(0)$ eigenvalues. We note that, although $\F_{\text{neq}}(\hat\rho_{\text{mb}}(0)) > \F_{\text{neq}}(\hat{\rho}(0))$, the former goes to equilibrium faster than the latter, implying an intersection in the cooling curves signature of the genuine quantum Mpemba effect as observed in Fig.~\ref{fig:fig3}.}
    \label{fig:fig1}
\end{figure}

Figure~\ref{fig:fig1} exemplifies the QME for a spin 1/2 system, depicting the theoretical cooling dynamics for different initial states. The plot shows that the transformed state, $\hat\rho_{\text{mb}}(0)$, despite having a higher initial free energy, reaches the equilibrium region (blue) faster than the state $\hat\rho(0)$, which started with a lower free energy. This faster relaxation is the core of the QME, and its dynamical signature is the crossing of their respective free-energy trajectories over time, as experimentally observed in Fig.~\ref{fig:fig3}. It is also worth noting that, although the \textit{passive state}~\cite{Pusz1978}, $\hat\rho_{\text{ps}}(0)$ (represented in Fig.~\ref{fig:fig1}) reaches equilibrium before other states, this does not constitute a QME since it begins with a lower free energy, and thus its faster approach to equilibrium is thermodynamically expected. \par

Given a state, $\hat\rho(0)$, a consistent method to achieve the unitary transformation that induces the QME is to maximize the non-equilibrium free energy. As aforementioned, a transformation that diagonalizes the state in the energy eigenbasis, followed by a population inversion, will not only result in a \textit{genuine} QME but also ensure that it is maximally pronounced.

\begin{figure}[t!]
\centering
\includegraphics[width=\linewidth]{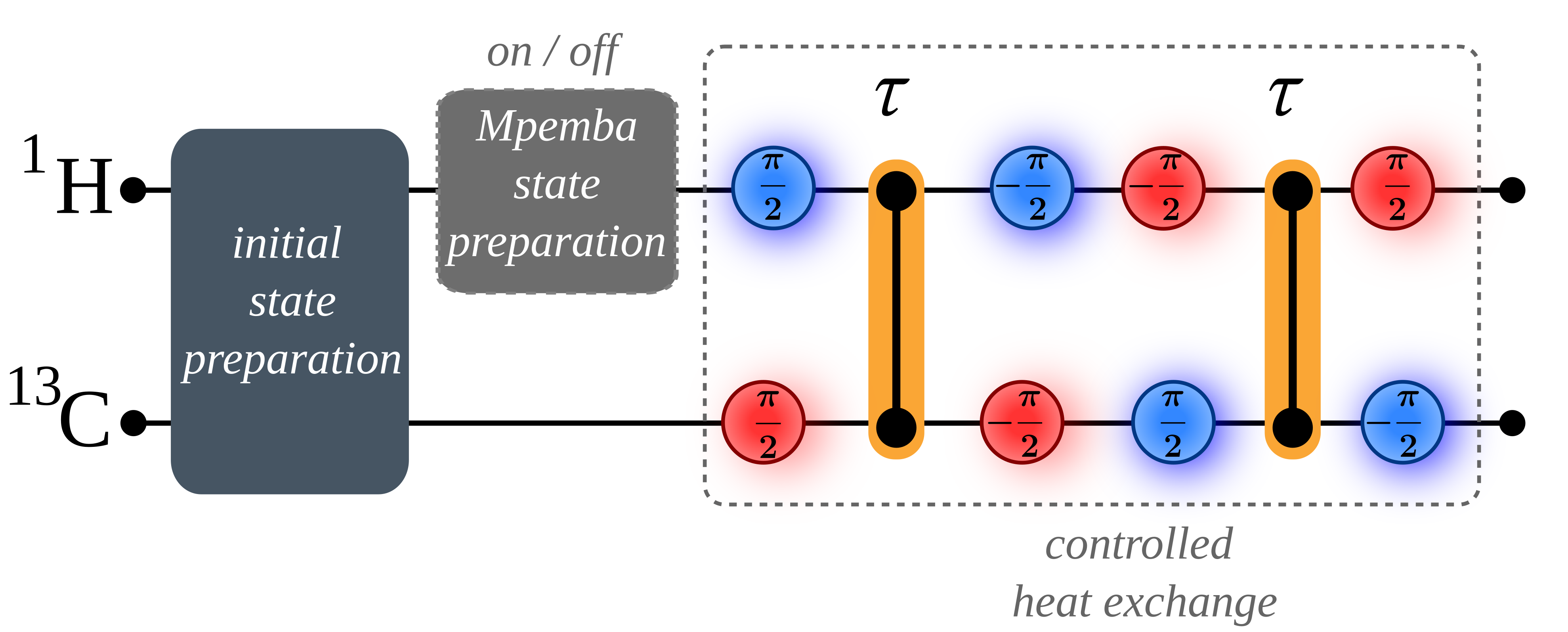}
\caption{A schematic depiction of the experimental protocols. In the first implementation, the heat exchange protocol is realized directly after the initial state preparation (without Mpemba state preparation). Conversely, the second implementation introduces an intermediate procedure that unitarily transforms the system to the QME state before heat exchange. Blue (red) circles denote rotations along the $x$ ($y$) direction, realized through rf-pulses. Yellow markings indicate periods of free evolution under scalar coupling with duration $\tau\in [0,(2J)^{-1}]$, as described by the interaction Hamiltonian $\hat H_{\text{int}}=2\pi\hbar J\hat\sigma_z\otimes\hat\sigma_z$, where $J=215.1$~Hz is the scalar coupling strength. See also the Supplemental Material.}
    \label{fig:fig2}
\end{figure}

\textit{Observing the Mpemba Effect}\textemdash NMR offers a level of preparation and control of spin systems that makes it a comprehensive platform for proof-of-principle tests in quantum thermodynamics~\cite{Batalhao2014,Micadei2019,Micadei2021,Mahesh_PRA22,DLu_PRL25}. Here, we consider a liquid sample of 99\% $^{13}$C-labeled CHCl$_{3}$ diluted in acetone-d6, where our working substance is encoded in the nuclear spin of $^1\text{H}$ atoms, while the nuclear spin of $^{13}\text{C}$ plays the role of a heat sink in the cooling protocol. All experiments were conducted using a $600$~MHz Bruker NMR spectrometer. By employing an appropriate sequence of radio-frequency (rf) pulses along with gradient field pulses~\cite{Peterson_2019,Vieira2023}, we prepare an effective thermal state $\hat{\rho}_0^{\text{eq}} = e^{-\beta\hat \H^{\text{eq}}}/Z_0$ on the $^{13}\text{C}$ with $\beta^{-1}=k_{B} T/ h = (9.22 \pm 0.07)\; \text{kHz}$,
as defined by the Hamiltonian $\hat \H^{\text{eq}} = -2\pi\hbar\nu_1\hat{\sigma_z}$, with the energy gap determined by the frequency $\nu_1 = 2\; \text{kHz}$. The rf offset fixes the Hamiltonian in the present implementation. Our goal is to examine the heat exchange between the $^{1}\text{H}$ working medium and the $^{13}\text{C}$ heatsink using the \textit{heat exchange} protocol depicted in Fig.~\ref{fig:fig2}, which fulfils the necessary conditions of a Davies map for the working medium reduced dynamics (more details in End Matter). We also initialize the $^{1}\text{H}$ nuclear spin into a non-equilibrium state equivalent to $\hat{\rho} (0) \approx \dyad{x_{+}}$, that
exhibits maximum coherence in the energy eigenbasis of $\hat \H^{\text{eq}}$. Here, we will consider two distinct scenarios. First, the heat exchange protocol is applied immediately following the state preparation \(\hat{\rho} (0)\), and we track the working medium evolution up to reaching thermal equilibrium. In the second scenario, before the heat exchange, a sequence of rf pulses is applied to transform the system state via the unitary operator $\hat{U}$, thereby promoting the QME. In this specific case, the system's state is transformed into a diagonal state with a population inversion in the energy eigenbasis, referred to as the QME state, $\hat{\rho}_\text{mb}(0)$, after which the heat exchange protocol is applied.

In both realized protocols, the non-equilibrium free energy variation, \(\Delta \mathcal{F}_{\text{neq}} = \mathcal{F}_{\text{neq}}(\hat\rho)-\F_{\text{eq}} = \beta^{-1}S_{KL}(\hat\rho(t)||\hat\rho^{eq})\), of the working medium is tracked along the heat exchange dynamics, as depicted in Fig.~\ref{fig:fig3} (where \(S_{KL}(\hat\rho||\hat\sigma)=\mathrm{Tr}\left[\hat\rho(\ln\hat\rho-\ln\hat\sigma)\right]\) is the Kullback-Leibler divergence). We observe a reasonable agreement between the theoretical simulation (solid curves) and the experimental results (circles) in  Fig.~\ref{fig:fig3}. When $\Delta \mathcal{F}_{\text{neq}} \sim  0$ the system approaches the Gibbs state. We observe in  Fig.~\ref{fig:fig3} that the protocol that cools directly, $\hat\rho (0)$, after state preparation (blue curve) approaches the equilibrium state slower than the protocol that first applies a suitable unitary transformation, $ \hat\rho_{\text{mb}}(0) = \hat U \hat{\rho} (0) \hat U^\dagger$, (red curve). The latter protocol starts in a state with higher non-equilibrium free energy (further from equilibrium). Additionally, the intersection of the thermalization curves is one of the signatures of the genuine QME~\cite{Mpemba2024}. As aforementioned, the mechanism behind the QME is the contribution of slow and fast Lindbladian eigenmodes along the heat exchange dynamics. Such contributions were experimentally observed and are presented in the End Matter.

\begin{figure}[t!]
\centering
\includegraphics[width=\linewidth]{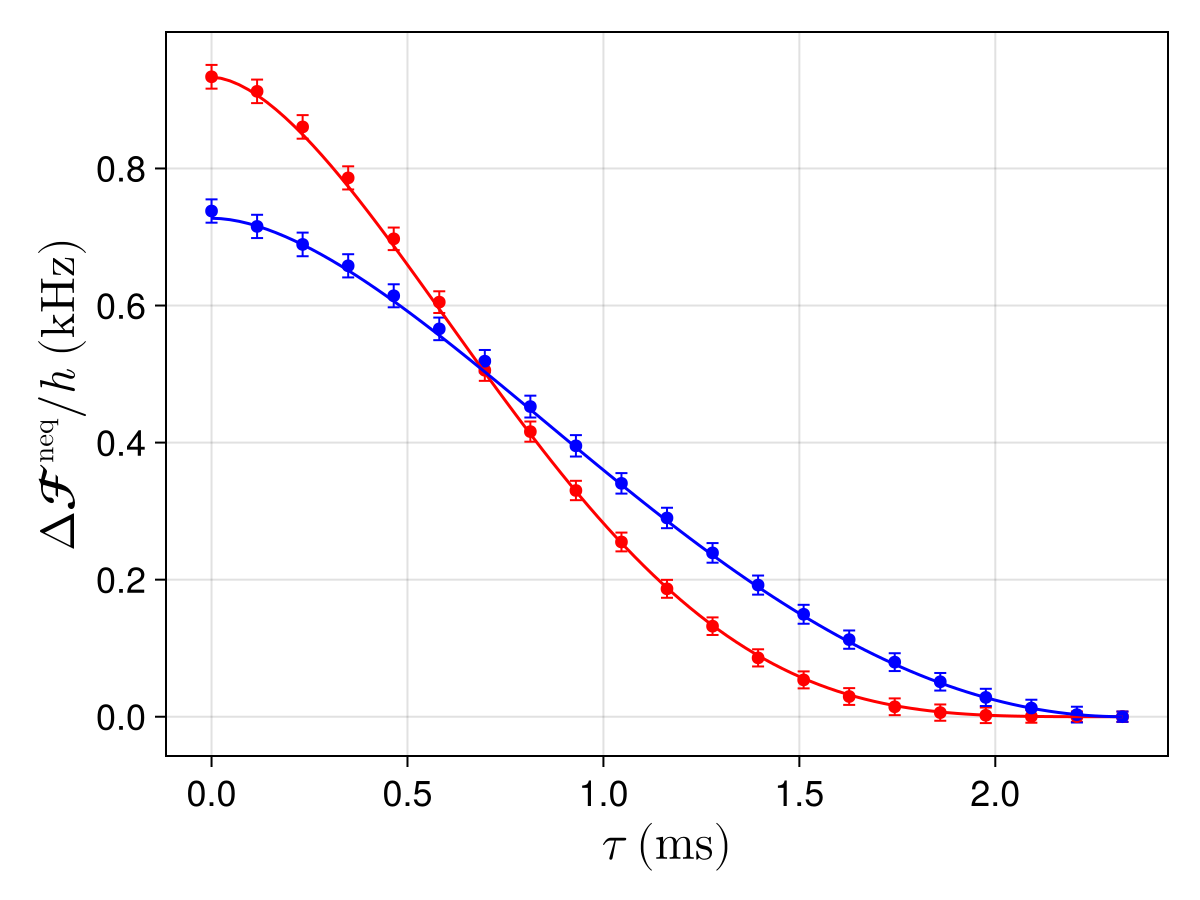}
\caption{Non-equilibrium free energy variation of the working medium along the heat exchange protocol. A comparison is shown between a system initially prepared in the state $\hat{\rho} (0)$ in blue, and in the QME state, $ \hat\rho_{\text{mb}}(0) = \hat U \hat{\rho} (0) \hat U^\dagger$, in red. Circles showcase experimental measurements, whereas solid lines depict the theoretical dynamics of the experimental states. The quantum Mpemba effect is evidenced by the characteristically faster dynamics towards equilibrium presented for the transformed state (in red), despite its initially higher non-equilibrium free energy. Error bars denote the propagated error estimated through Monte Carlo iterations. The error bars that are smaller than the size of the symbols are not shown.}
    \label{fig:fig3}
\end{figure}

\textit{Application in a quantum Otto refrigerator}\textemdash~ Furthermore, we investigate the use of the QME in the quantum Otto refrigerator to speed up the cycle, thereby gaining a cooling power advantage. 
To provide a proof-of-concept, we start with the working medium in the equilibrium state with the cold environment, $\hat\rho_0^{\text{eq,c}}$, defined by the Hamiltonian $\hat\H_0^{\text{eq}} = -2\pi\hbar\nu_0\hat\sigma_x$, at temperature $k_BT_c/ h = (307 \pm 3)\; \text{Hz}$. Subsequently, we implement a
linear energy gap expansion, realized experimentally by a modulated rf-pulse, resulting in the Hamiltonian~\cite{Quench_2015}, during the interval $t \in [0,\tau_1]$, 
\begin{equation}
    \hat\H^{\text{exp}} = -2\pi\hbar\,\nu(t,\tau_1)\,\hat\sigma_x\;,
\end{equation}
where $\nu (t,\tau_1) = \nu_0 (1-t/\tau_1) + \nu_1 t/\tau_1$ is a linear modulation of the rf field intensity with an appropriate phase, $\tau_1= 100\;\mu\text{s}$ is the final time of the gap expansion dynamics with $\nu_0 = 1 \; \text{kHz}$ ($\nu_1 = 2 \; \text{kHz}$). At this point, the QME unitary transformation can be applied or not (see supplementary Fig.~S1). Next, the cooling protocol (depicted in Fig.~\ref{fig:fig2}) with the auxiliary qubit prepared with $k_{B} T_h/ h = (4.77 \pm 0.07)\; \text{kHz}$ which concludes at time $\tau_2$. The third stroke is then performed until $\tau_3$, lasting for a duration equal to that of the first stroke. During this stroke, the working medium energy gap is compressed to its original value ($\nu_1 \to \nu_0$) through the time-reversed dynamics given by $\hat\H^{\text{comp}} = \hat\Theta\hat\H^{\text{exp}}\hat\Theta^{-1}$, where \(\hat\Theta\) is the time reversal operator. Finally, the cycle can be closed by a complete thermalization with the cold environment, which reaches the state $\hat\rho_0^{\text{eq,c}}$ at time $\tau_4 = (2J)^{-1}$.\\

Let us introduce the average heat extracted from the cold environment, which is given by  $\expval{Q_c} = \Tr[\hat\H_0^{\text{eq}}(\tau_1)\hat\rho^{\text{eq,c}}] - \Tr[\hat\H_0^{\text{eq}}\hat\rho_{\tau_3}]$. So, the cooling power can be defined as $P = \expval{Q_c}/\tau_f$, where $\tau_f$ is the total time required for one cycle realization. For the sake of simplicity, we introduce the duration of the $i$-th stroke as $\Delta\tau_i = \tau_i - \tau_{i-1}$. In this setting we consider a variable cooling stroke duration $ \Delta\tau_2 \in [0,(2J)^{-1}]$ such that the full cycle execution time is $\tau_4 = \bar{\tau} + \Delta\tau_2$, where $\bar{\tau} \equiv \Delta\tau_1 + \Delta\tau_3 + \Delta\tau_4 \approx 4.65 \; \text{ms}$ corresponds to the time due to strokes with a fixed duration. We consider the trace distance $\D(\hat\rho(t_2),\hat\rho_0^{\text{eq,h}})=\text{Tr}|\hat\rho(t_2) -\hat\rho_0^{\text{eq,h}}|/2$ as a distinguishability measure~\cite{Nielsen2000} between the working-medium instantaneous state during the cooling stroke, $\hat\rho(t_2)$, and the fixed point $\hat\rho_0^{\text{eq,h}}$, where $t_2 \in [\tau_1,\tau_2]$. Hence, a cooling power advantage (due to the QME) is obtained if there exists a final cooling time $\tau_2\mapsto \D(\hat\rho_{mb}(\tau_2),\hat\rho_0^{\text{eq,h}}) < \D(\hat\rho(\tau_2),\hat\rho_0^{\text{eq,h}}) = \delta$, where $\delta$ is a trace distance threshold fixed by the endpoint $\tau_2$ chosen in the implementation of the second stroke.   
\par
\begin{figure}[h!]
    \centering
    \includegraphics[width=\linewidth]{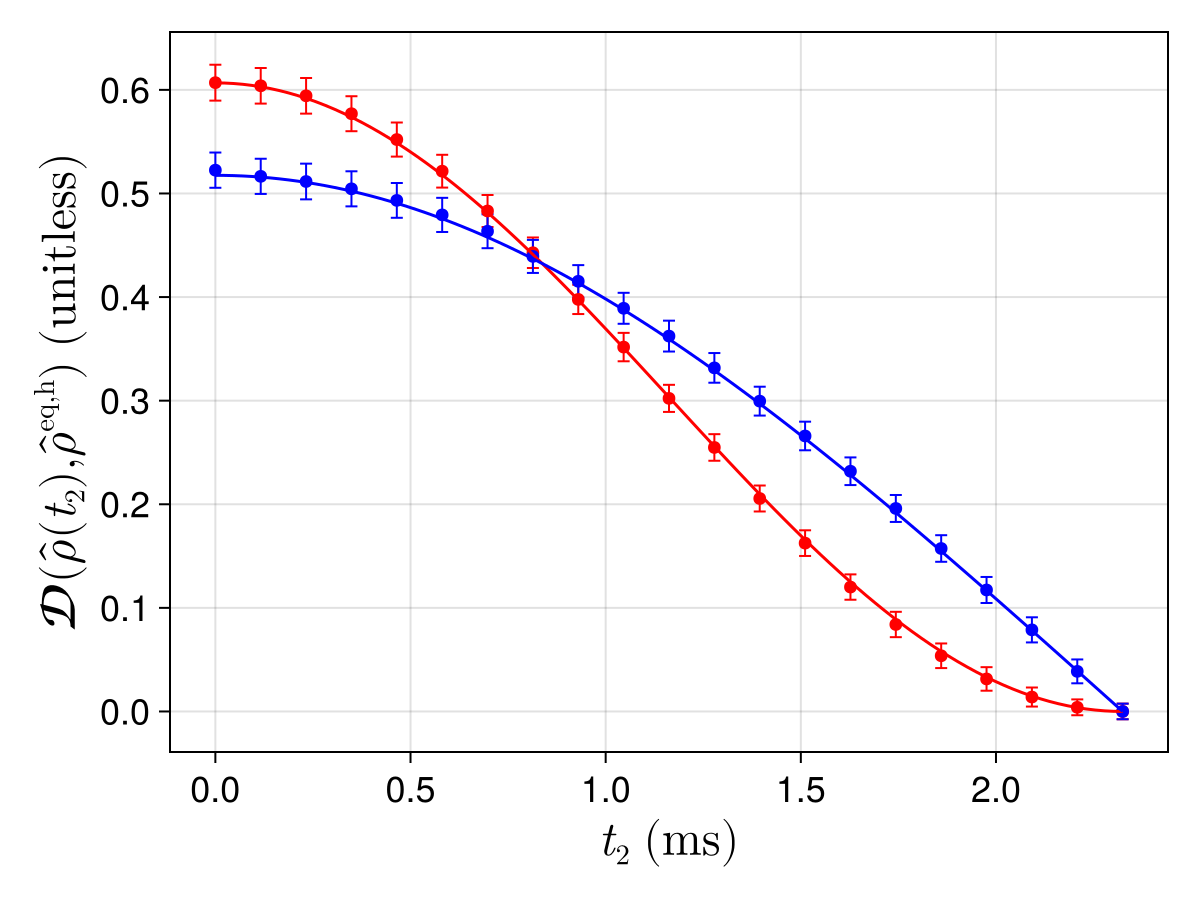}
    \caption{Trace distance between the evolved state \(\hat\rho(t_2)\) and the target thermal state \(\hat\rho^{\text{eq,h}}\) during thermalization. The red (blue) curve represents the theoretical cooling of the system prepared with (without) the optional QME preparation step, whereas the markers correspond to experimental measurements. We observe that preparing the QME state leads to a characteristically faster approach to equilibrium. The quantity $\D(\hat\rho(t_2),\hat\rho_0^{\text{eq,h}})$ measures the distinguishability between the instantaneous state of the system, $\hat\rho(t_2)$, and the asymptotic state of the dynamics, $\hat\rho_0^{\text{eq,h}}$. Its values are employed as a threshold, $\delta$, in the application of the QME to the quantum Otto refrigerator seen in Fig.~\ref{fig:fig5}. Error bars denote the propagated error estimated through Monte Carlo iterations. The error bars that are smaller than the size of the symbols are not shown.}
    \label{fig:fig4}
\end{figure}
\begin{figure}[h!]
    \centering
    \includegraphics[width=\linewidth]{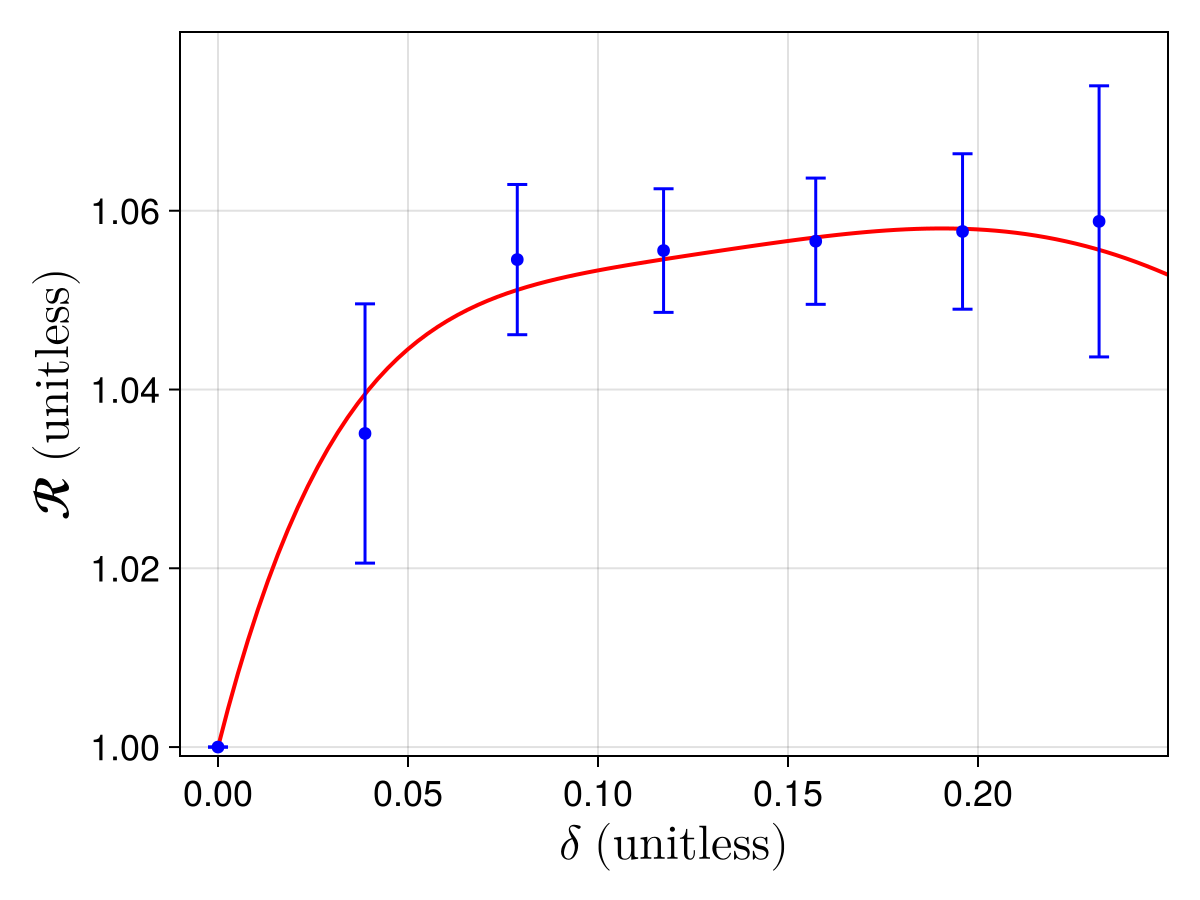}
    \caption{Experimental power gain ratio $\mathcal{R}$, seen in Eq.~\eqref{eq:pw}, as a function of the threshold \(\delta\) depicted in Fig.~\ref{fig:fig4}. Blue points represent experimental measurements, while the solid red curve shows the trend line. The cooling power advantage peaks at the point of maximum separation between the thermalization curves in Fig.~\ref{fig:fig4}. Error bars denote the propagated error estimated through Monte Carlo iterations.}
    \label{fig:fig5}
\end{figure}

Hence, the advantage in cooling power can be quantified in terms of the second stroke's duration $\Delta\tau_2^0$ ($\Delta\tau_2^{mb}$), such that the system initially in the state $\hat\rho (0)$ ($\hat\rho_{mb}(0)$) reaches a fixed threshold $\delta$ according to the ratio
\begin{equation}\label{eq:pw}
    \mathcal{R}(\delta) = \frac{P_{mb}(\delta)}{P_0(\delta)} = \frac{\bar{\tau} + \Delta\tau_2^0(\delta)}{\bar{\tau} + \Delta\tau_2^{mb}(\delta)}\;,
\end{equation}
where we note that $\mathcal{R}(\delta)\geq 1$ as a consequence of the $\delta$ definition. The figure of merit introduced in Eq.~\eqref{eq:pw} is very versatile for highlighting the QME advantage in the refrigerator cycle. In Fig.~\ref{fig:fig5}, we observe that the power ratio of Eq.~\eqref{eq:pw} initially equals one (for $t_2=2.13$~ms) and increases as the separation between the trace distance curves in Fig.~\ref{fig:fig4} grows, peaking at the point of maximum separation ($t_2\approx 1.74$~ms) and subsequently decreasing back to one as the curves converge to cross in Fig.~\ref{fig:fig4} at $t_2 \approx 0.87$~ms. In the present implementation, the power gain ($\mathcal{R}(\delta)$) reaches about 6\%, demonstrating an interesting performance enhancement from using the QME in thermal tasks compared to an otherwise identical cycle without the QME.

The observed cooling power advantage relies on preparing the state $\hat\rho_{mb}(0)$, which has a minimum intrinsic cost due to the difference in non-equilibrium free energy between the initial and transformed states. This cost inherently reduces the efficiency of the QME-based quantum thermal refrigerator. Our result thus integrates the QME into the power-efficiency trade-off paradigm~\cite{EfPw1,EfPw2,EfPw3,EfPw4}, revealing it as a new mechanism for balancing power against efficiency—a central consideration for optimizing quantum devices.

\textit{Conclusions}\textemdash~We have observed the genuine QME in the cooling protocol of a spin-1/2 system, which has been recently described theoretically in the context of Davies maps~\cite{Mpemba2024}.  By exploring the interplay between coherence in the energy eigenbasis and the dynamics of superoperator eigenmodes, our experiment highlights the connection between thermodynamics, quantum mechanics, and information theory. Furthermore, we leveraged the QME to increase the cooling power of the quantum Otto refrigerator by accelerating thermalization, thus demonstrating new possibilities for enhancing quantum thermal machines. As a future research direction, it would be interesting to investigate the efficiency-power trade-off introduced by QME across different scenarios and thermal applications. 

\textit{Note added}\textemdash~Recently, during the final stages of completion of our manuscript, we became aware of an independent work~\cite{Mahesh2025} that reports an observation of the genuine QME in NMR employing different methods.

\textit{Acknowledgments}\textemdash B.P.S. acknowledges the São Paulo Research Foundation (FAPESP) Grant No.2024/12957-2 and Conselho Nacional de Pesquisa (CNPq) Grant No.140240/2026-8  for financial support. J.L.D.O. acknowledges the Coordenação de Aperfeiçoamento de Pessoal de Nível Superior, Brazil (CAPES), Finance Code 001. Y.A.C.A acknowledges CNPq Grant No.189533/2025-0. C.H.S.V. acknowledges the São Paulo Research Foundation (FAPESP) Grant No. 2023/13362-0 and Grant No. 2025/14546-2 for financial support and the Southern University of Science and Technology (SUSTech) for providing the workspace. K.Z. acknowledges CNPq Grant No.305665/2025-1. R.M.S. acknowledges financial support from CNPq, CAPES, and FAPESP. The authors also acknowledge FAPESP EMU project No. 2022/11550-0.

\bibliography{references2}

\onecolumngrid
\vspace{\baselineskip}
\noindent\rule{\textwidth}{1pt}
\begin{center}
    \large\textbf{End Matter}
\end{center}
\vspace{\baselineskip}
\twocolumngrid
\newpage
\clearpage

\textit{Experimental Identification of Davies Map}\textemdash~The experimental characterization of a general quantum dynamics can be performed through quantum process tomography (QPT), which maps the complete quantum channel associated with a physical process. This is achieved by determining the Choi-Jamiolkowski matrix~\cite{Chuang1997, Nielsen2000}, $\xi$, corresponding to the quantum channel, $\mathcal{E}(\hat\rho)=\sum_{i,j=0}^3\xi_{i,j} \hat\Xi_i \hat\rho \hat\Xi^\dagger_j$, where we have chosen the following operator basis: 
\begin{align*}
\hat\Xi\in \{ &(\hat{\mathds{1}} + \hat\sigma_z)/2, \,(\hat\sigma_x+i\hat\sigma_y)/2, \\ 
&(\hat\sigma_x-i\hat\sigma_y)/2, \, (\hat{\mathds{1}} - \hat\sigma_z)/2 \}.
\end{align*}

To experimentally characterize the heat exchange protocol, we performed a stroboscopic QPT varying the protocol time duration, $\tau$. We have prepared the $^1\text{H}$ nuclear spin in a set of states
of a mutually unbiased basis (MUB)~\cite{Chuang1997, Nielsen2000}, and subjected each one to the heat exchange protocol for a chosen time duration. The output density matrix is reconstructed using QST and numerical post processing of the acquired data enables the determination of the process (Choi) matrix, $\xi$. An example of the experimental process matrix obtained is shown in Fig.~6 (more details and results can be found in the Supplemental Material). We observe that the experimental process matrix is very similar the theoretical one (i.e. the generalized amplitude damping) with worst case fidelity of $98.3\%$.
\begin{figure}[h!]
    \centering
    \includegraphics[width=\linewidth]{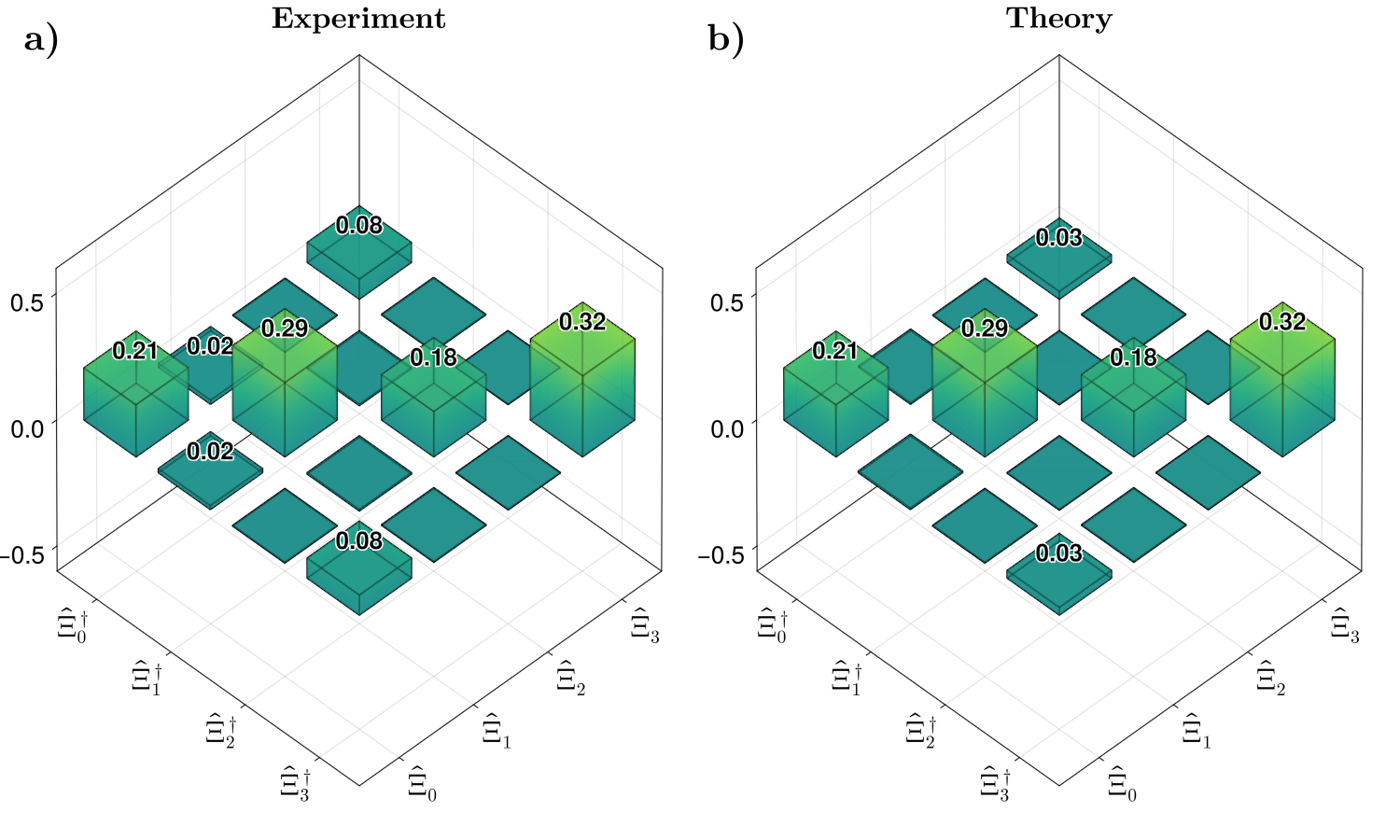}
    \caption{Experimental (a) and theoretical (b) real process (Choi) matrix for a particular cooling duration ($\tau \sim 2.04$ ms). For clarity, numerical labels are added only to elements $\xi_{ij}$ such that $|\Re{\xi_{ij}}| \geq 0.010$. Also, the imaginary part of each element is negligible (under experimental precision) and hence suppressed from the figure. Analogue figures for all times where QPT was performed can be found in the Supplemental Material.}
    \label{fig:fig6}
\end{figure}
From the Choi matrix representation of the physical map implemented, the Davies map properties can be directly observed as follows (see Supplemental Material for more details):

\textit{Commutativity and Separability conditions}\textemdash Davies maps require that the unitary and dissipative components of the dynamics are separable and that they commute with each other. These properties impose strong constraints on the structure of the Choi matrix associated with a Davies map. Specifically, in the chosen operator basis $\{\hat\Xi_i\}$, the Choi matrix assumes an almost-diagonal form with vanishing off-diagonal elements except for $\xi_{14}=\xi_{41}^*$, which describes the damping of quantum coherences in the density operator. In fact, these properties can be observed in the experimental data, as in Fig.~\ref{fig:fig6} (see also Eq.~(S17) and Fig.~S4 in the Supplemental Material). Indeed, in the series of  QPT performed, the values of the off-diagonal elements are zero within the measurement precision, except for $\xi_{14}=\xi_{41}^*$.

\begin{figure}[h!]
    \centering
    \includegraphics[width=0.8\linewidth]{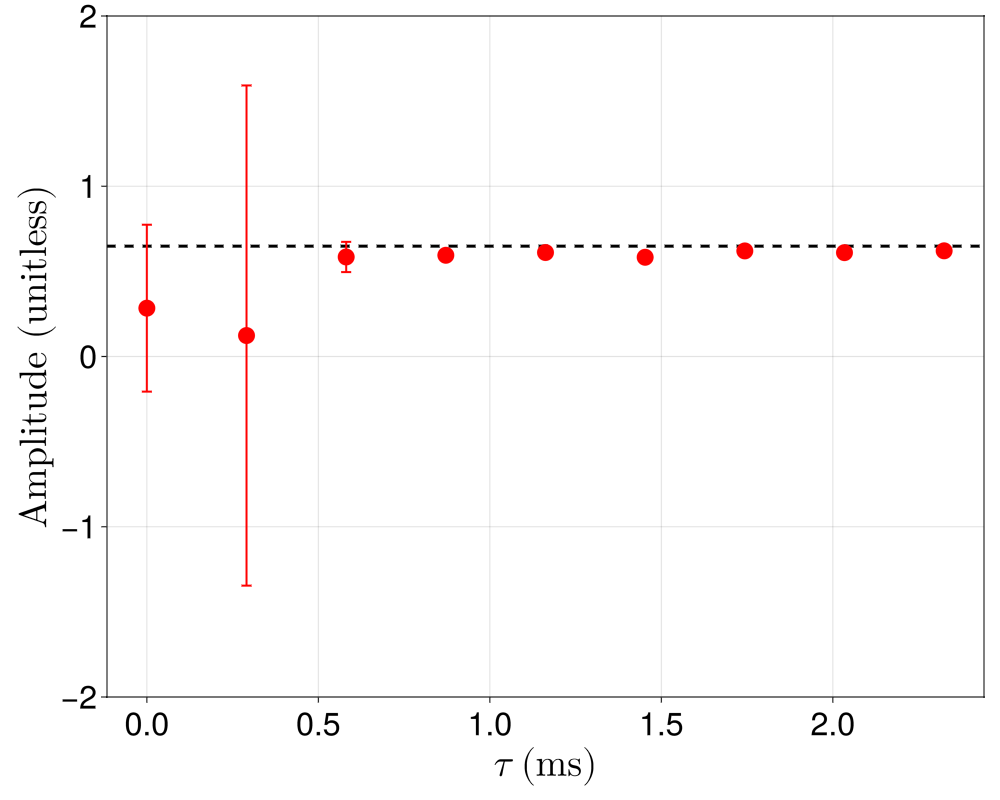}
    \caption{Experimental verification of the Quantum Detailed Balance condition in the heating protocol. Following Eq.~\eqref{eq::qdb}, the dashed line represents the theoretical Boltzmann factor, $e^{-2\beta h\nu_1}$, whereas the dots represent the ratio $\xi_{33}/\xi_{22}$ obtained from the experimental processes (Choi) matrices.}
    \label{fig:fig7}
\end{figure}

\textit{Quantum Detailed Balance condition (QDB)}\textemdash~The QDB condition relates the ratio of the excitation ($\Gamma_{\uparrow}$) and relaxation rates ($\Gamma_{\downarrow}$) with the thermal Boltzmann factor, $e^{-2\beta h\nu_1}$. In the chosen operator basis $\{\hat\Xi_i\}$, the excitation/relaxation ratio can be expressed in terms of the process matrix elements as (see Supplemental Material): 
\begin{equation}\label{eq::qdb}
\frac{\Gamma_\uparrow}{\Gamma_\downarrow}
=
\frac{\xi_{33}}{\xi_{22}}
=e^{-2\beta h\nu_1}.
\end{equation}
In fact, from the QPT experimental data, we observe that such condition is fulfilled along the heat exchange protocol as depicted in Fig.~\ref{fig:fig7}, where we plot $\xi_{33}/\xi_{22}$ for different heat exchange time durations, which should lie on the line defined by $e^{-2\beta h\nu_1}$ according to Eq.~(5). 

\textit{Dynamic Contribution of Lindblad eigenmodes}\textemdash~
In Ref.~\cite{Mpemba2024} and in the main text, it is argued that the underlying mechanism behind the speed-up to equilibrium and the Mpemba effect is the contribution of the slowest eigenmode to the dynamics. From the experimental Choi matrix obtained through QPT, one can obtain the Kraus operator decomposition ($\hat K_j(\tau)$) associated with the (discrete) quantum channel, at a specific heat exchange duration $\tau$~\cite{gilchrist2011}, in the Liouville space,
\begin{equation}
    \xi (\tau) = \sum_{j=1}^4 |\hat K_j(\tau)\rrangle\llangle \hat K_j(\tau)| \;.
\end{equation}

\begin{figure} 
    \includegraphics[width= \linewidth]{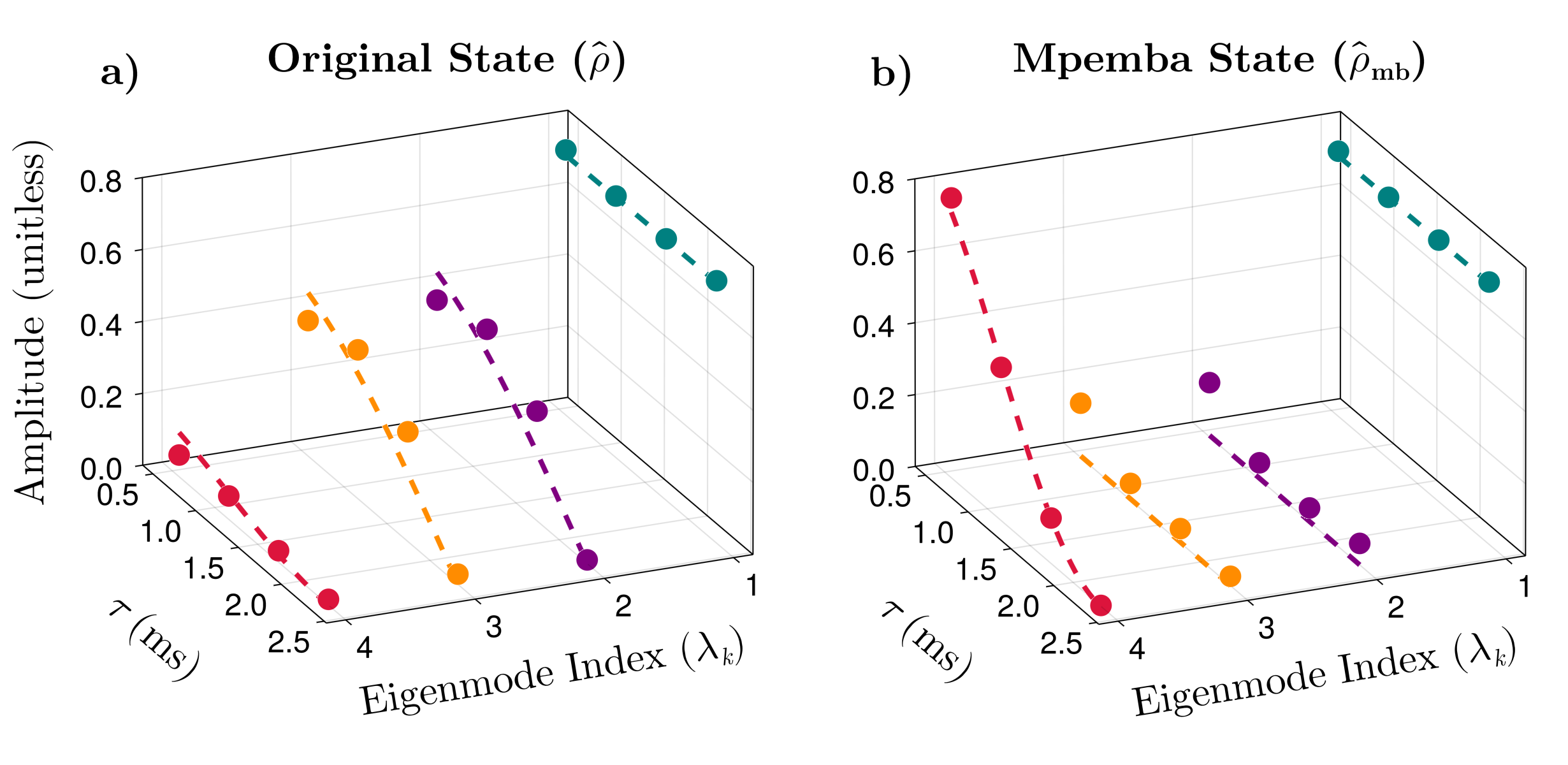}
    \caption{Projection of the instantaneous experimental state of the system onto each (experimental) Lindbladian eigenmode $\alpha_k=\llangle \hat\xi_k|\hat\rho(0)\rrangle$ as defined in Eq.~(\ref{eq::lindblad_sol}). The dashed line represents the theoretical factor $\alpha_k$ and the dots represents the experimental results, where the eigemodes are in the increasing order, mode 1 (steady state) $\lambda_1 \approx 0.0$~kHz, mode 2 and 3 (slow-decaying quantum coherences) $\Re(\lambda_{2,3})\approx 0.2$~kHz, and mode 4 (fast-decaying population relaxation) $\Re(\lambda_4) \approx 0.4$~kHz. a) For original initial state, $\hat\rho$, and b) for the transformed (Mpemba) state, $\hat\rho_{\text{mb}}$.}
    \label{fig:fig8}
\end{figure}

In the context of Markovian dynamics, the semi-group generator \(\L\) in Liouville space of a map $\mathcal{E}$ (that admits $\mathcal{E}^{-1}$) possessing a Kraus decomposition, can be derived from the relation~\cite{Gyamfi_2020}:
\begin{equation}\label{eq:sup_lindbladian}
    \L = \frac{1}{t}\ln \sum_{j=1}^4 \hat K_j(t) \otimes \hat K_j(t)^\dagger\;.
\end{equation}
Hence, the Lindbladian super-operator $\L$ can be experimentally determined from the QPT. In conjunction with the experimental quantum states at specific times of the heat exchange, we obtain the projection of the instantaneous state of the system onto each Lindbladian eigenmode $\alpha_k=\llangle \hat\xi_k|\hat\rho(0)\rrangle$, associated with the decay rate $\Re(\lambda_k)$ as defined in Eq.~(\ref{eq::lindblad_sol}). This result, presented in Fig.~\ref{fig:fig8}, uncovers the mechanism underlying the Mpemba effect. We observe, in Fig.~\ref{fig:fig8}, that the original state exhibits a larger contribution of the slowest decaying coherence modes (modes 2 and 3) in its cooling dynamics, whereas the transformed (Mpemba) state exhibits a small contribution of these coherence modes, and a very large contribution of the fastest decaying population mode (mode 4) to its dynamics.

\end{document}


\preprint{APS/123-QED}

\title{Supplemental Material: Experimental observation and application of the genuine Quantum Mpemba Effect}

\maketitle

Here, we provide additional details regarding the open dynamics in Liouville space, the implementation of a Davies map in a Nuclear Magnetic Resonance setup, the quantum Otto refrigerator proof-of-concept, additional figures, and error analysis.

\section{Open dynamics and Lindbladian superoperator}
The Lindbladian superoperator in Eq. (1) of the main text has the form
\begin{multline}\label{eq::lindblad_super}
        \mathbf{\L} = -\frac{i}{\hbar} [\![\hat\H, \mathds{\hat I}_d]\!]\; + \\ +\sum_{k=1}^{d^2 - 1} \gamma_k( \hat A_k \otimes \hat A_k^* - \frac{1}{2} [\![ \hat A_k^\dagger \hat A_k, \mathds{\hat I}_d ]\!]_+ )\;,
\end{multline}
in terms of the Hamiltonian, the so-called jump operators, $\hat A_k$, $\gamma_k$ are the decay rates related to their respective transitions provoked by $\hat A_k$,$[\![\hat A,\hat B]\!] = \hat A\otimes \hat B - \hat A \otimes \hat B^{\text{T}}$ and $[\![\hat A,\hat B]\!]_{+} = \hat A\otimes \hat B + \hat A \otimes \hat B^{\text{T}}$ are the generalized commutators. In this notation, operators with a hat are in Schrodinger space. Furthermore, their left and right eigenvectors form a biorthonormal basis~\cite{Mingati2019} and thus satisfy the completeness relation
\begin{equation}\label{eq::complet}
    \bm{\mathds{\hat 1}}_{d^2} = \sum_{k=1}^{d^2}|\hat \zeta_k\rrangle\llangle \hat\xi_k|\;.
\end{equation}
Therefore, by integrating Eq.~(1) of the main text and inserting the relation~\eqref{eq::complet}, one arrives at Eq.~(2) of the main text
\begin{align} 
    |\hat \rho(t)\rrangle &=e^{t\L}|\hat\rho(0)\rrangle \label{eq:sup_dynamics}\\
    &=e^{t\L}\bm{\mathds{\hat1}}_{d^2}|\hat\rho(0)\rrangle\\
    &=\sum_{k=1}^{d^2}e^{t\L}|\hat \zeta_k\rrangle\llangle \hat\xi_k||\hat\rho(0)\rrangle\\ 
    &=\sum_{k=1}^{d^2}e^{t\lambda_k}|\hat \zeta_k\rrangle\llangle \hat\xi_k||\hat\rho(0)\rrangle \;.
\end{align}
Additionally, from this eigendecomposition of the Lindbladian superoperator, we see that the steady state of the dynamics $|\hat \rho(t\rightarrow\infty)\rrangle$ is either the right eigenstate $|\hat\zeta_k\rrangle$ corresponding to $\lambda_k = 0$, or, if $\lambda_k = 0$ has a multiplicity greater than one, a linear combination of those.

\section{Experimental implementation of the heat exchange}
The heat exchange protocol depicted in Fig.~2 of the main text has the effect of a non-linear map applied locally to the $^1$H qubit, which can be represented as
\begin{equation}\label{eq:Kraus}
\mathcal{E}(\hat\rho) = \sum_{j=1}^4 \hat K_j \hat\rho(0) \hat K_j^\dagger
\end{equation}
with the corresponding Kraus operators $\hat K_j$, in the computational basis, of the map given by 
\begin{equation}
\hat K_1 = \sqrt{1-p_1^{\text{aux}}}
\begin{bmatrix}
1 & 0 \\
0 & \cos(\pi J \tau)
\end{bmatrix},
\end{equation}

\begin{equation}
\hat K_2 = \sqrt{1-p_1^{\text{aux}}}
\begin{bmatrix}
0 & \sin(\pi J \tau) \\
0 & 0
\end{bmatrix},
\end{equation}

\begin{equation}
\hat K_3 = \sqrt{p_1^{\text{aux}}}
\begin{bmatrix}
\cos(\pi J \tau) & 0 \\
0 & 1
\end{bmatrix},
\end{equation}

\begin{equation}
\hat K_4 = \sqrt{p_1^{\text{aux}}}
\begin{bmatrix}
0 & 0 \\
-\sin(\pi J \tau) & 0
\end{bmatrix},
\end{equation}
where $p_1^{\text{aux}} \in [0, 1/2]$ is the excited state population of the auxiliary ($^{13}$C) qubit at time $\tau$. During the experiment, the delay time is varied within the interval  $\tau \in [0, (2J)^{-1}]$. Under these conditions, the transformation in Eq.~\eqref{eq:Kraus} is equivalent to a generalized amplitude damping map~\cite{Srikanth2008}, which corresponds to the Davies map describing the thermalization of a single qubit~\cite{petruccione2002}. Consequently, from the local reference frame of the $^{1}$H qubit, the experimental protocol is indistinguishable from that of the system undergoing thermalization via a Davies map. \\
\textcolor{blue}{In the context of Markovian dynamics}, the semi-group generator \(\L\) in Liouville space of an invertible map $\mathcal{E}$ \textcolor{blue}{possessing} a Kraus decomposition, such as in Eq.~\eqref{eq:Kraus}, can be derived from the relation~\cite{Gyamfi_2020}
\begin{equation}\label{eq:sup_lindbladian}
    \mathbf{\L} = \frac{1}{t}\ln \sum_{j=1}^4 \hat K_j(t) \otimes \hat K_j(t)^\dagger\;,
\end{equation}
which allows for direct computation of the Lindbladian spectrum and its eigenvectors. \\

For instance, in our experimental setup, the aforementioned map can be implemented by the pulse sequence denoted in Fig.~2. Additionally, considering the set of initial states defined by $\hat\rho_\theta(0) = \hat R_y(\theta)\hat\rho (0)\hat R_y(-\theta)$, where $\hat{\rho} (0) \approx \dyad{x_{+}}$, $|x_{\pm}\rangle$ are the $\hat\sigma_x$ Pauli operator eigenstates (with eigenvalues $\pm 1$) and $ \hat R_y(\theta)$ is the usual SU2 rotation.  By inserting \eqref{eq:sup_lindbladian} into \eqref{eq:sup_dynamics} and vectorizing this set of initial states, it is possible to obtain the theoretical dynamics of the non-equilibrium free energy shown in Fig.~1 of the main text.\\

\section{Refrigerator setup}

\begin{figure}[h]
    \centering
    \includegraphics[width=\linewidth]{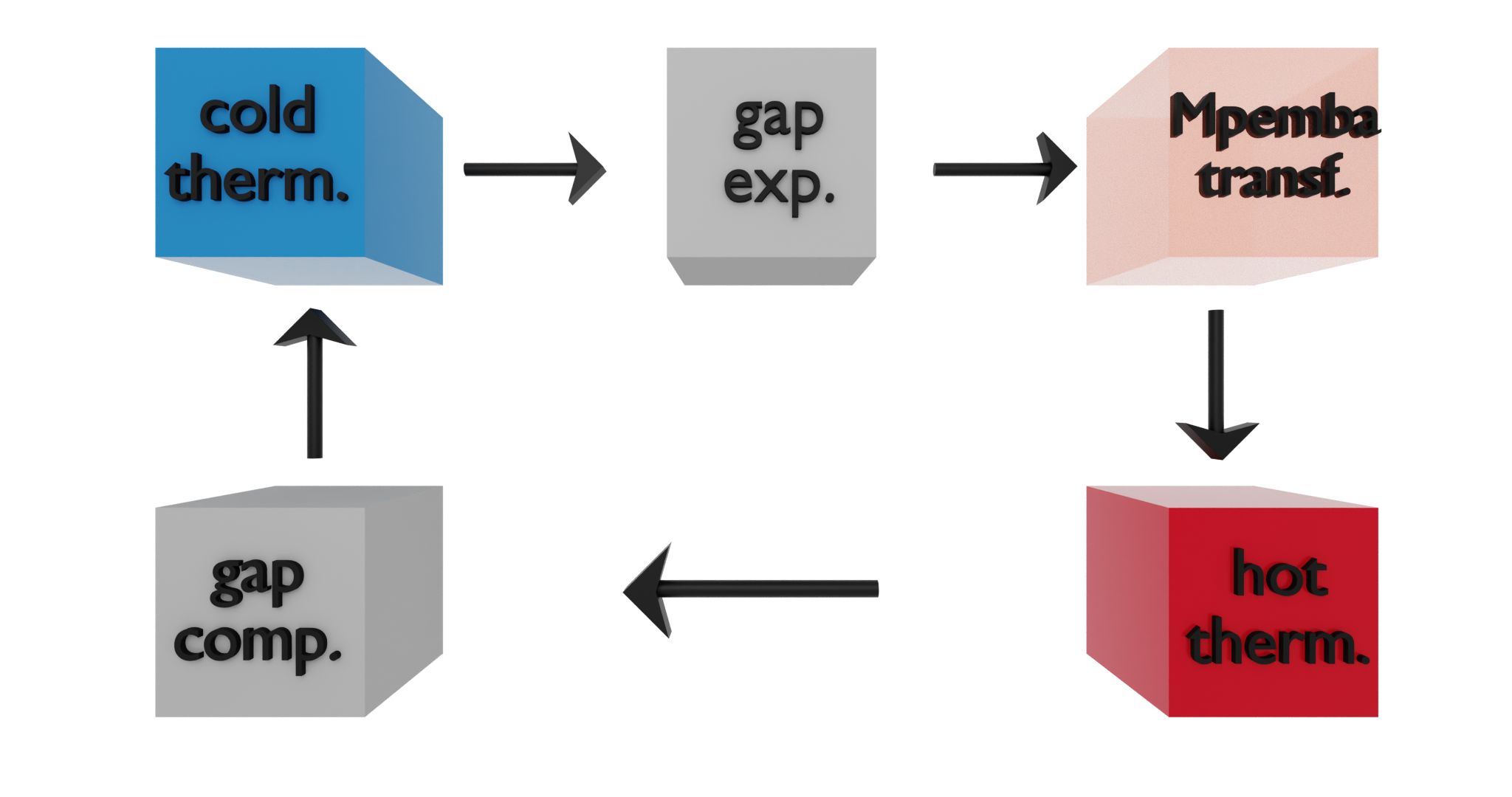}
    \caption{Schematic representation of the processes in the quantum Otto refrigerator proof-of-concept, highlighting the transformation to the QME state as a possible additional step.}
    \label{fig:sup1}
\end{figure}

A visual representation of the refrigeration procedure described in the main text is shown in Fig.~\ref{fig:sup1}, where an additional transformation before the cooling process can be added to the conventional protocol. Therefore, the whole cycle is composed of strokes one to four, each with a \textcolor{blue}{final time} $\tau_i, \; i=1,2,3,4$. Starting from the equilibrium state with the cold environment, the cycle strokes are: energy gap expansion, cooling protocol, energy gap compression and heating protocol, with the possible addition of a 5th step to exploit the QME before the cooling step. 

Hence, the conventional realization can be described by the time-dependent Hamiltonian 
\begin{equation}
    \H(t) = \begin{cases}
            \hat\H^{\text{exp}}(t) = -2\pi\hbar\,\nu(t,\tau_1)\,\hat\sigma_x\;, \; &  t \in [0,\tau_1] \\
        \hat\H^{\text{eq}} = -2\pi\hbar\nu_1\hat\sigma_z \;,\; &  t \in [\tau_1,\tau_2]\\
        \hat\H^{\text{comp}}(t) = \hat\Theta\hat\H^{\text{exp}}\hat\Theta^{-1} \;, \; &  t \in [\tau_2,\tau_3] \\
        \H^{\text{cold}} = -2\pi\hbar\nu_0\,\hat\sigma_x \;, \; &  t \in [\tau_3,\tau_f] \\

    \end{cases} \;,
\end{equation}\\
following the main text definitions. Here it's worth noting that, in our proof-of-concept experiment, the compression and expansion strokes can be realized by a single time-dependent rf pulse, with duration on the order of $\mu \text{s}$. In contrast, the cooling protocol is implemented via the pulse sequence depicted in Fig.~2 of the main text, which has a duration on the order of ms. Additionally, the preparation of the QME state, $\hat\rho_{\text{mb}}$, before the cooling process can also be realized by rf pulses of $\mu\text{s}$ order of magnitude.

\section{Additional figure supplementary to Fig.~1 of the main text}
Below follows a supplementary figure to Fig.~1 from the main text, indicating the Mpemba crossing of the thermalization curves for the numerical simulation provided in the main text. See also Fig.~S3.

\begin{figure}[h]
    \centering
    \includegraphics[width=\linewidth]{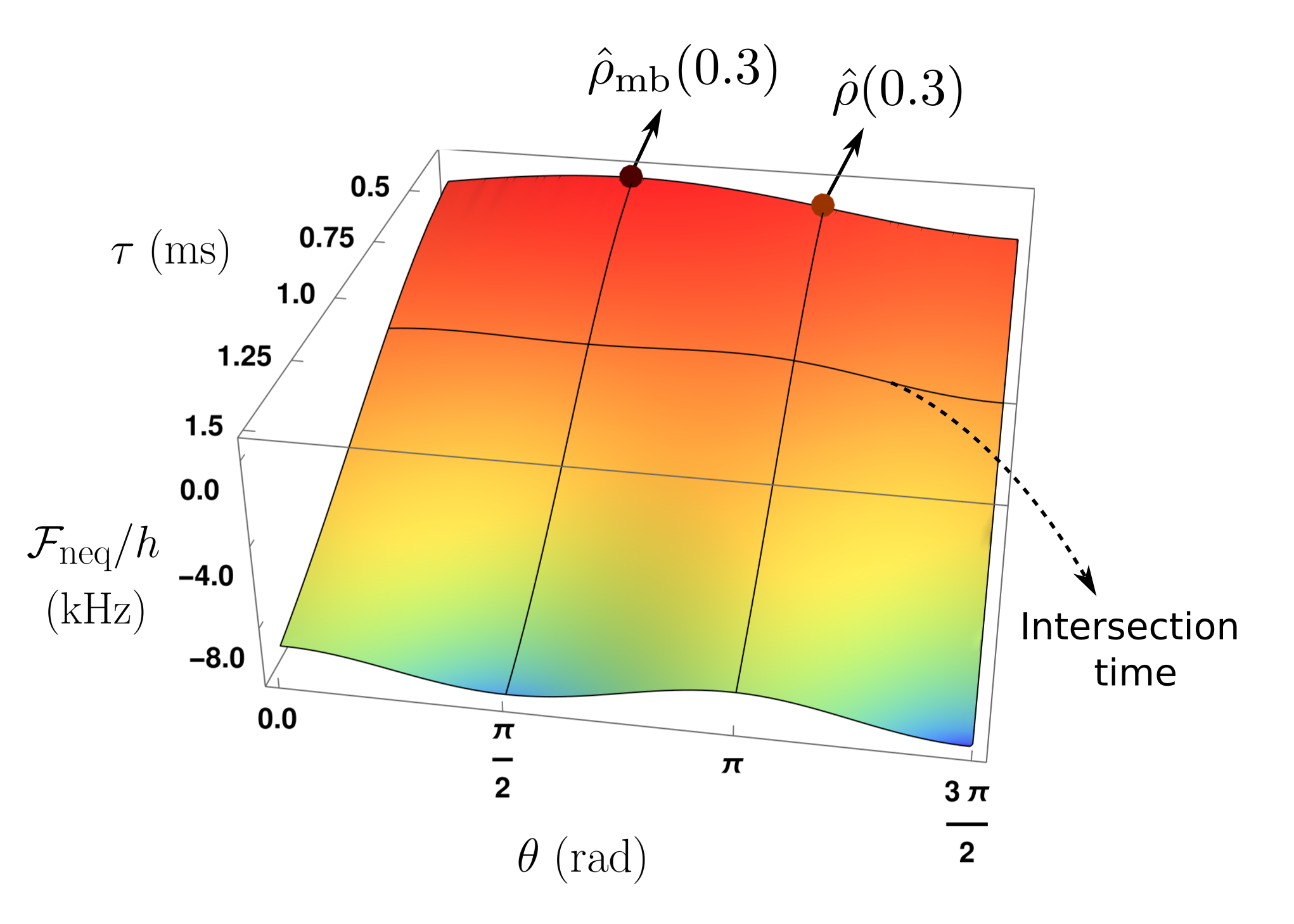}
    \caption{Schematics of the quantum Mpemba effect supplementing Fig.~1 of the main text. The final portion of relaxation dynamics is presented, highlighting the time, for which the non-equilibrium free energy of the initial states $\hat{\rho} (0) \approx \dyad{x_{+}}$ and $\hat\rho_{\text{mb}}(0) = \hat R_y(\pi/2)\hat\rho (0)\hat R_y(-\pi/2)$ coincide.}
    \label{fig:sup2}
\end{figure}

\begin{figure}[h]
    \centering
    \includegraphics[width=\linewidth]{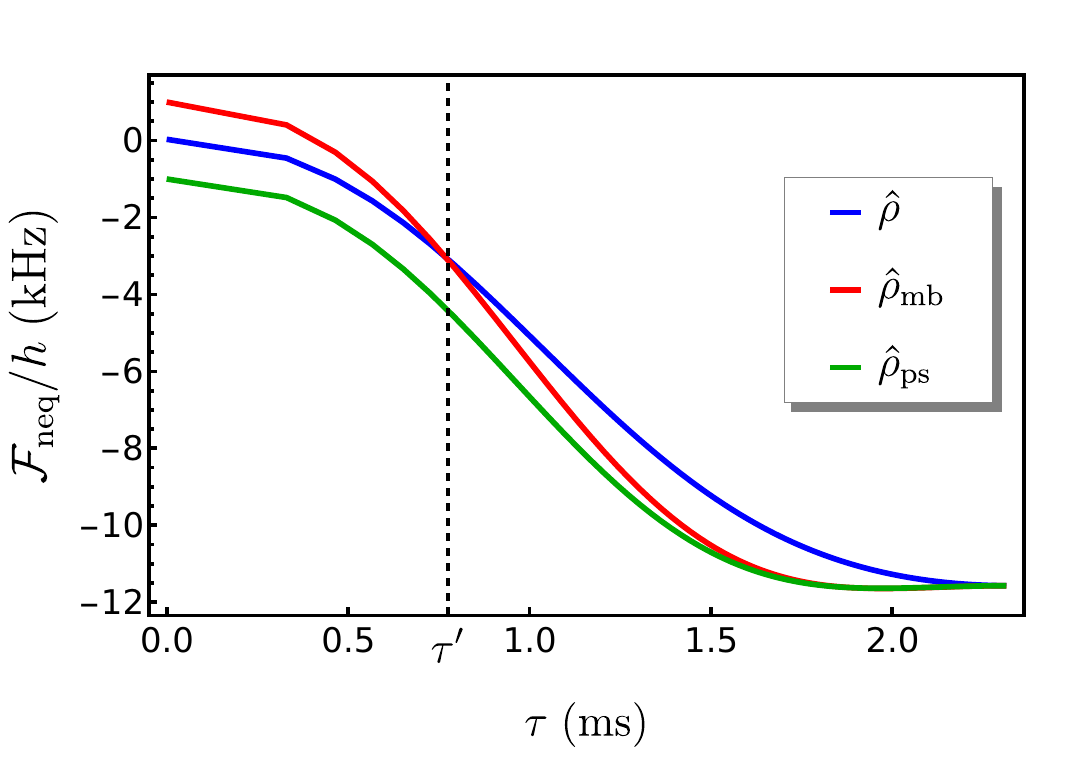}
    \caption{Projection of the relaxation dynamics shown in Fig.~\ref{fig:sup2} onto the plane perpendicular to the \(\theta\) axis. By suppressing the \(\theta\) coordinate, the crossing point at \(\tau'\approx0.775~\text{ms}\) becomes clearly visible, corresponding to the point where the non-equilibrium free energies of the initial states \(\hat\rho(0)\) and \(\hat\rho_{\text{mb}}(0)\) coincide.}
    \label{fig:sup3}
\end{figure}

\section{Characterization and Structural Conditions for Davies Map}

A particularly important class of open-system dynamics is given by \textit{Davies maps}, which describe Markovian thermalization under weak coupling to a thermal reservoir. The primary goal of this section is to establish the exact mathematical constraints that a general completely positive trace-preserving (CPTP) map, $\mathcal{E}$, acting on a single qubit with Hamiltonian $\hat{\mathcal{H}}_0$, must satisfy to be rigorously classified as a Davies map.

Generally, a single-qubit quantum channel qualifies as a Davies map if and only if it simultaneously fulfills two fundamental physical conditions~\cite{Roga2010}:
\begin{enumerate}
    \item \textbf{Decoupling of Populations and Coherences:} The dynamical evolution must completely suppress any population-to-coherence or coherence-to-population conversion.
    \item \textbf{Quantum Detailed Balance:} The map must satisfy the mathematical condition of detailed balance, ensuring thermodynamic consistency in the transition rates between energy levels.
\end{enumerate}


To understand how these physical principles constrain the mathematical structure of the map, it is highly instructive to first analyze $\mathcal{E}$ in the Pauli operator basis, $\{\hat{\mathds{1}},\hat\sigma_x,\hat\sigma_y,\hat\sigma_z\}$. In this representation, a generic qubit state is mapped to a Bloch vector $(1, r_x, r_y, r_z)^T$. The populations are encoded in the longitudinal component, $r_z=\rho_{00}-\rho_{11}$, while the coherences are encoded in the transverse components, $r_x=\rho_{01}+\rho_{10}$ and $r_y=i(\rho_{10}-\rho_{01})$.

The first condition (Decoupling of Populations and Coherences) dictates that the longitudinal and transverse sectors must evolve independently. Therefore, the matrix representation of any Davies map in the Pauli operator basis, $\xi'_{\mathrm{Davies}}$, is strictly forbidden from containing cross-terms that mix these sectors, assuming the sparse form~\cite{Nielsen2000}:
\begin{equation}
    \xi'_{\mathrm{Davies}}=
    \begin{pmatrix}
        1 & 0 & 0 & 0 \\
        0 & \eta_\perp & 0 & 0 \\
        0 & 0 & \eta_\perp & 0 \\
        \kappa & 0 & 0 & \eta_z
    \end{pmatrix}.
    \label{eq:Davies_Pauli_Basis}
\end{equation}
Under this map, the Bloch vector transforms cleanly as
\begin{equation}
    (r_x, r_y, r_z) \longrightarrow (r_x\eta_\perp, r_y\eta_\perp, \kappa+r_z\eta_z),
\end{equation}
demonstrating that coherences attenuate via $\eta_\perp$  independently of the population shifts dictated by $\kappa$ and $\eta_z$. 

To connect this dynamical picture to the standard Choi-Jamiołkowski process matrix representation, $\mathcal{E}(\hat\rho)=\sum_{i,j=1}^4\xi_{i,j} \hat\Xi_i \hat\rho \hat\Xi^\dagger_j$, we perform a change of basis in the operator space. We define our sorted transition operators $\hat\Xi_k$ explicitly in terms of the Pauli matrices as:
\begin{align}
    \hat\Xi_1 &= |0\rangle\langle0| = \frac{\hat{\mathds{1}} + \hat\sigma_z}{2}, \nonumber\\
    \hat\Xi_2 &= |0\rangle\langle1| = \frac{\hat\sigma_x+i\hat\sigma_y}{2}, \nonumber\\
    \hat\Xi_3 &= |1\rangle\langle0| = \frac{\hat\sigma_x-i\hat\sigma_y}{2}, \nonumber\\
    \hat\Xi_4 &= |1\rangle\langle1| = \frac{\hat{\mathds{1}} - \hat\sigma_z}{2}.
\end{align}
Crucially, the population operators ($\hat\Xi_1, \hat\Xi_4$) are linear combinations exclusively of the longitudinal Pauli operators ($\hat{\mathds{1}}, \hat\sigma_z$), whereas the coherence operators ($\hat\Xi_2, \hat\Xi_3$) are constructed strictly from the transverse ones ($\hat\sigma_x, \hat\sigma_y$). 

Although the evolution matrix in the Pauli basis (Eq.~(\ref{eq:Davies_Pauli_Basis})) is not strictly diagonal due to the thermal drive term $\kappa$, it completely preserves the separation between the longitudinal sector ($\hat{\mathds{1}}, \hat\sigma_z$) and the transverse sector ($\hat\sigma_x, \hat\sigma_y$). This strict decoupling is perfectly inherited during the linear change of basis. Consequently, any cross-term $\xi_{ij}$ in the new process matrix that connects a population operator to a coherence operator is algebraically forced to vanish. This structural inheritance guarantees that the process matrix $\xi_{\mathrm{Davies}}$ assumes a highly sparse form, where only population-population and coherence-coherence transitions are non-vanishing:
\begin{equation}
    \xi_{\mathrm{Davies}} =
    \begin{pmatrix}
        \xi_{11} & 0 & 0 & \xi_{14} \\
        0 & \xi_{22} & 0 & 0 \\
        0 & 0 & \xi_{33} & 0 \\
        \xi_{41} & 0 & 0 & \xi_{44}
    \end{pmatrix}.
    \label{eq:davies_choi_structure}
\end{equation}
Any single-qubit CPTP map that does not conform to this inherited matrix structure cannot physically represent a Davies thermalization process.

A prime example of a physical quantum process that strictly conforms to these derived criteria is the Generalized Amplitude Damping (GAD) channel~\cite{Khatri2020}. The GAD channel perfectly instantiates this sparse Davies structure, with its process matrix in the operator basis  $\{\hat\Xi_k\}$ explicitly given by
\begin{equation}
    \xi_{\mathrm{GAD}} = 
    \begin{pmatrix} 1-p_1\gamma & 0 & 0 & \sqrt{1-\gamma} \\ 0 & p_1\gamma & 0 & 0 \\ 0 & 0 & (1-p_1)\gamma & 0 \\ \sqrt{1-\gamma} & 0 & 0 & 1-(1-p_1)\gamma 
    \end{pmatrix}.\label{eq:gad_choi}
\end{equation}
By defining $\gamma$ as the system-reservoir interaction strength and $p_1$ as the excited-state population at thermal equilibrium, this exact correspondence establishes the GAD channel as the standard paradigm for single-qubit Davies maps.

Having established the sparse matricial structure imposed by the decoupling condition, we now turn to the second fundamental requirement: quantum detailed balance. Physically, the longitudinal dynamics—which are now strictly isolated within the population sector of the map—describe incoherent excitation and relaxation processes between the energy eigenstates,
\[
|0\rangle
\leftrightarrow
|1\rangle,
\]
Letting $\Gamma_\uparrow$ and $\Gamma_\downarrow$ denote the excitation and relaxation rates, respectively, the quantum detailed balance condition~\cite{petruccione2002} imposes a strict thermodynamic constraint on their ratio:
\begin{equation}
\frac{\Gamma_\uparrow}{\Gamma_\downarrow}
=
e^{-\beta (\epsilon_1-\epsilon_0)},
\label{eq:detailed_balance}
\end{equation}
where $\epsilon_0$ and $\epsilon_1$ are the energies of the ground and excited states of $\hat{\mathcal{H}}_0$. Rather than allowing arbitrary population transfers, this condition ensures that the microscopic transition rates do not drive the system arbitrarily, but strictly respect the energetic constraints dictated by the surrounding thermal environment. Ultimately, this thermodynamic consistency in the transition rates perfectly balances the excitation and relaxation flows, naturally determining the asymptotic populations of the system and guaranteeing the emergence of the Gibbs state.

\section{Experimental Identification of the Heat Exchange Protocol as a Davies Map}

To experimentally verify whether a reconstructed quantum channel is compatible with Davies thermalization, the physical process is completely characterized through quantum process tomography (QPT). In our implementation, the system consists of a single qubit governed by the Hamiltonian
\begin{equation}
    \hat{\mathcal{H}}^{\text{eq}} = -2\pi\hbar\nu_1\hat\sigma_z.
\end{equation}

The tomographic reconstruction employs a complete set of single-qubit input states, specifically the six states associated with the mutually unbiased Pauli bases:
\begin{equation}
    \left\{
    |0\rangle,
    |1\rangle,
    \frac{|0\rangle+|1\rangle}{\sqrt2},
    \frac{|0\rangle-|1\rangle}{\sqrt2},
    \frac{|0\rangle+i|1\rangle}{\sqrt2},
    \frac{|0\rangle-i|1\rangle}{\sqrt2}
    \right\},
    \label{eq:mub_states}
\end{equation}
which correspond to antipodal points along the three Cartesian axes of the Bloch sphere. 

From the experimentally measured input-output relations, the Choi matrix, $\xi$, associated with the quantum channel is obtained. The process reconstruction is performed via maximum-likelihood estimation, which ensures that the resulting channel is a physically admissible completely positive trace-preserving (CPTP) map.

Let $\xi_{\text{exp}}$ denote the experimentally reconstructed process matrix in the standard computational operator basis, $\{\Xi_k\}$. To rigorously identify the implemented protocol as a Davies thermalization map, we systematically evaluate $\xi_{\text{exp}}$ against the two structural constraints established in the previous section, as well as their thermodynamic consequence.

The first constraint, the decoupling of populations and coherences, imposes a highly sparse structure on the process matrix (see Eq.~\ref{eq:davies_choi_structure}). Experimentally, this condition is verified by inspecting the magnitude of the mixing elements in $\xi_{\text{exp}}$. For the process to be compatible with a Davies map, all matrix elements $\xi_{ij}$ linking the population subspace to the coherence subspace must vanish up to statistical fluctuations and experimental noise; that is, $\xi_{ij} \approx 0$ for all $(i,j) \notin \{(1,1), (2,2), (3,3), (4,4), (1,4), (4,1)\}$.

Provided the decoupling condition holds, we proceed to verify the second requirement: quantum detailed balance. Crucially, the heat exchange protocol realized in our experiment is designed to be structurally equivalent to a GAD channel~\cite{schnepper2026optimizations}. Because of this equivalence, the non-vanishing elements of the experimental matrix $\Xi_{\text{exp}}$ can be directly mapped to the theoretical GAD parameters. Under this framework, the diagonal elements of Eq.~\eqref{eq:gad_choi} directly encode the transition probabilities associated with thermal excitation and relaxation processes. Specifically, the matrix element corresponding to the excitation from the ground state to the excited state is given by
\begin{equation}
    P_{0\rightarrow1} = p_1\gamma \equiv \xi_{33},
    \label{eq:excitation_probability}
\end{equation}
while the element corresponding to the relaxation from the excited state to the ground state is given by
\begin{equation}
    P_{1\rightarrow0} = (1-p_1)\gamma \equiv \xi_{22},
    \label{eq:relaxation_probability}
\end{equation}
up to the normalization convention adopted for the process matrix.

The ratio between the effective excitation and relaxation rates is therefore directly extracted from the experimental data as
\begin{equation}
    \frac{\Gamma_\uparrow}{\Gamma_\downarrow} = \frac{P_{0\rightarrow1}}{P_{1\rightarrow0}} = \frac{\xi_{33}}{\xi_{22}}.
    \label{eq:experimental_ratio}
\end{equation}
Consequently, the detailed-balance condition can be experimentally verified by comparing this measured ratio, $\xi_{33}/\xi_{22}$, with the theoretical thermal Boltzmann factor $e^{-2\beta h\nu_1}$. An agreement between these quantities provides direct evidence that the reconstructed quantum channel satisfies thermodynamic consistency, and it was presented in Fig.~7 in the End Matter of the main text.

\begin{figure}[htbp] 
    \centering
        \centering
        \includegraphics[width=\linewidth]{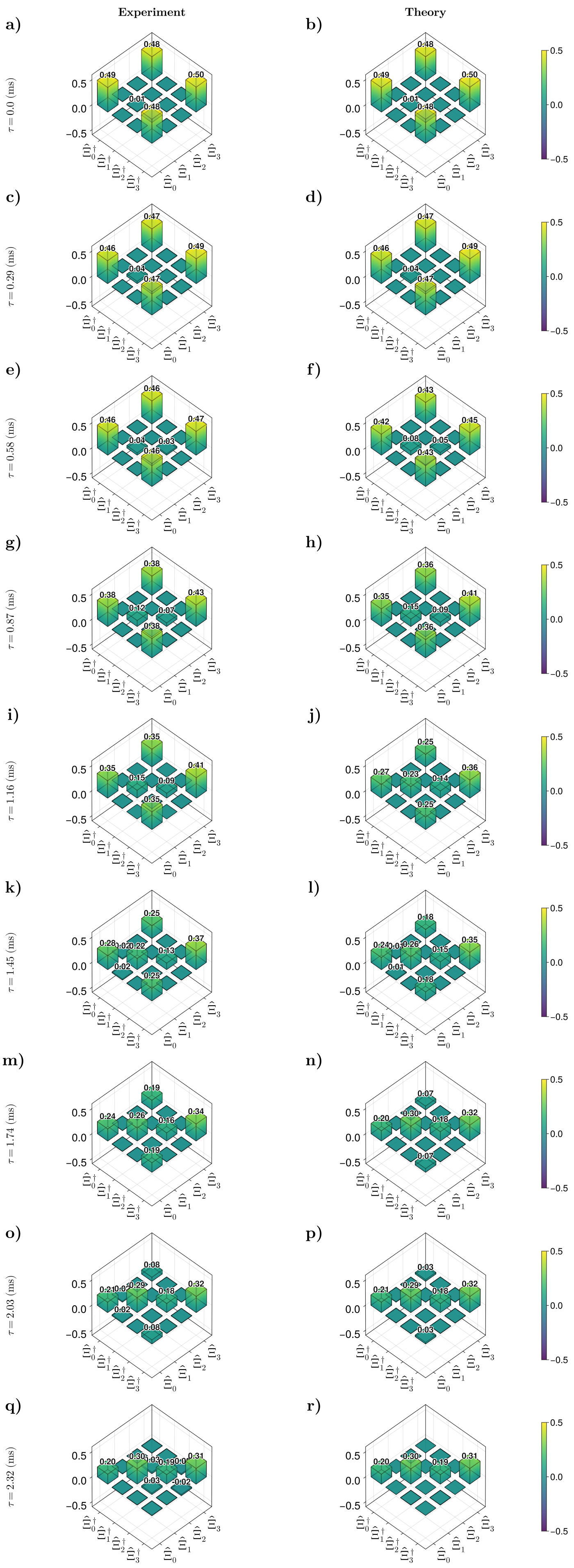}
        \caption{Comparison between experimentally reconstructed process (Choi) matrices -- left column figures-- and theoretical prediction -- right column figures, for multiple heat exchange durations.}
        \label{fig:sup4}
\end{figure}

It is worth mentioning that the Choi matrix represents a discrete channel. Hence, for a dynamical map such as the thermalization map considered in the main text, QPT is performed for a set of discrete time points. For a set of heat exchange times, the experimental and predicted process (Choi) matrices are showcased in Fig.~\ref{fig:sup4}, where numerical labels are added only to elements $\xi_{ij}$ such that $|\Re{\xi_{ij}}| \geq 0.010$ for improved clarity. Also, the imaginary part of each element is negligible (under experimental precision) and hence suppressed from the figure.

\section{Contribution of the slowest Lindbladian eigenmode to the dynamics}

To showcase the mechanism behind the speed-up to equilibrium of the Mpemba effect, we show the dynamical contribution of the slowest Lindbladian mode in the heat exchange protocol. From the experimental Choi matrix obtained through QPT, one can obtain the instantaneous Kraus operator decomposition ($\hat K_j(\tau)$) associated with the (discrete) quantum channel, at a specific heat exchange duration $\tau$~\cite{gilchrist2011}, in the Liouville space,
\begin{equation}
    \xi (\tau) = \sum_{j=1}^4 |\hat K_j(\tau)\rrangle\llangle \hat K_j(\tau)| \;.
\end{equation}
This relation, in conjunction with Eq.~\eqref{eq:sup_lindbladian}, allows us to experimentally determine the Lindbladian super-operator $\L$ from the quantum process tomography. With the experimental quantum states and by diagonalizing the Lindbladian at specific times of the heat exchange, we obtain the contribution of each Lindbladian eigenmode $\alpha_k=\llangle \hat\xi_k|\hat\rho(0)\rrangle$ to its dynamics. 

Because the implemented dynamics conform to a Davies map, these eigenmodes physically decouple into transverse (coherence) and longitudinal (population) sectors. This can be verified by mapping the eigenvectors back to $2 \times 2$ matrices. Each eigenmode is associated with a decay rate $\Re(\lambda_k)$, in increasing order: mode 1 represents the thermal steady state ($\lambda_1 \approx 0.0$~kHz); modes 2 and 3 are characterized by purely off-diagonal elements and thus govern the slow decay of quantum coherences ($\Re(\lambda_{2,3})\approx 0.2$~kHz); finally, mode 4 is purely diagonal, thereby dictating the fast relaxation of populations ($\Re(\lambda_4) \approx 0.4$~kHz). 

\begin{figure}[h]
        \centering
        \includegraphics[width=\linewidth]{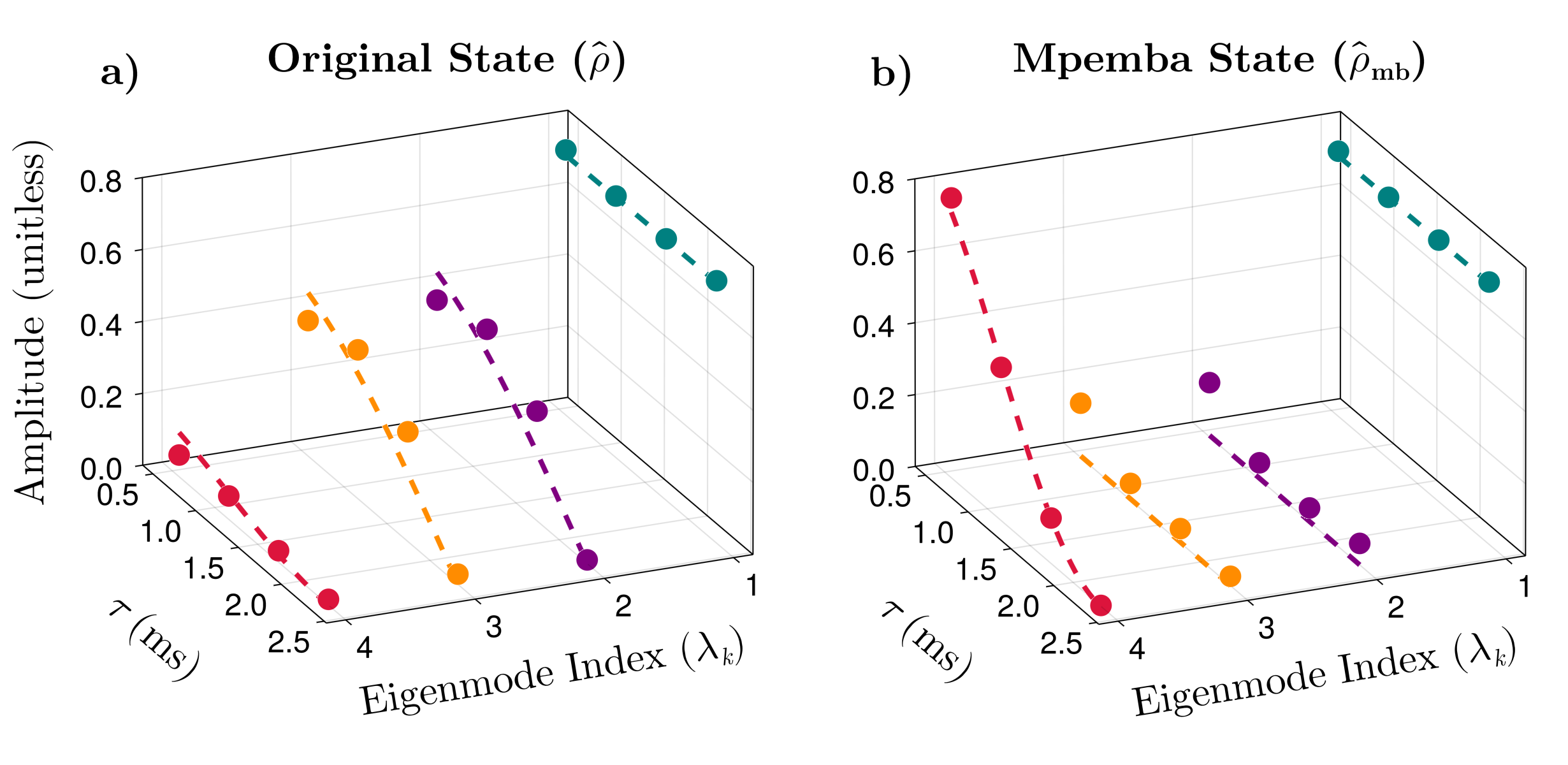}
        \caption{Experimental versus theoretical amplitudes for all four eigenmodes of the Liouvillian. The Mpemba state ($\hat{\rho}_{mb}$) and the original state ($\hat{\rho}$) are projected onto the left eigenbasis as a function of the relaxation time $\tau$. The solid lines with markers indicate the experimental data, while the dashed lines represent the continuous theoretical map.}
        \label{fig:sup5}
\end{figure}

This spectral separation, presented in Fig.~\ref{fig:sup5}, uncovers the mechanism underlying the Mpemba effect. We observe, in Fig.~\ref{fig:sup5}, that the original state exhibits a large contribution of the slow-decaying coherence modes (modes 2 and 3) in its cooling dynamics. Conversely, the transformed (Mpemba) state exhibits a small contribution of these slow coherence modes, and a very large contribution of the fast-decaying population mode (mode 4) to its dynamics. By bypassing the slow coherence bottleneck, this characteristic explicitly demonstrates why the Mpemba state speeds up to equilibrium in the heat exchange dynamics.

\begin{figure*}[t] 
    \centering
        \centering
        \includegraphics[width=0.8\linewidth]{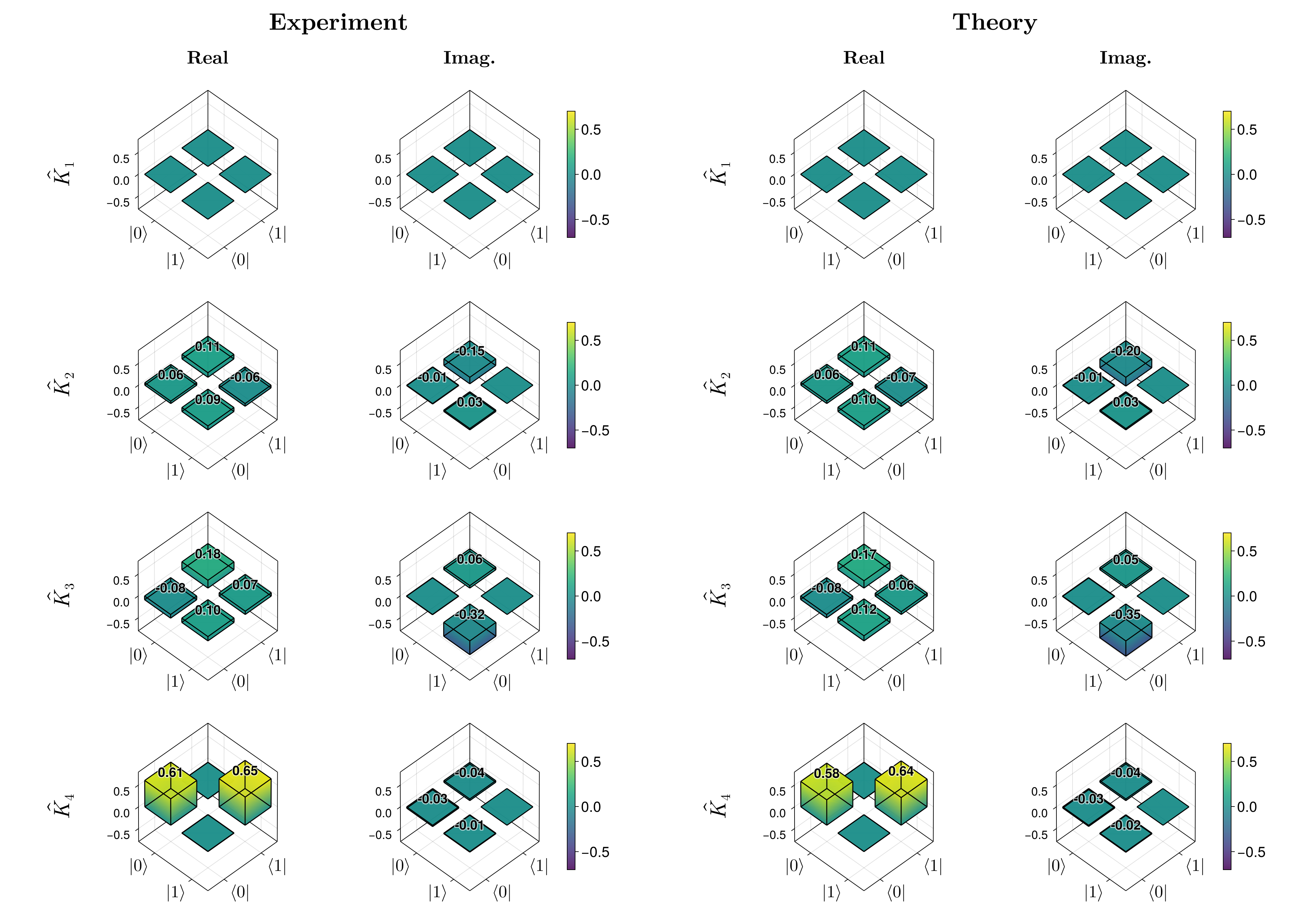}
        \caption{Comparison between Kraus operators of the experimentally reconstructed process, left-column figures,  and theoretical prediction, right-column figures,  for $\tau = 0.87$ ms.}
        \label{fig:sup6}
\end{figure*}

For completeness, Fig.~\ref{fig:sup6} displays a comparison of the Kraus operators experimentally obtained with this procedure and the theoretical prediction, for a particular heat exchange duration. In the figure, elements smaller than $0.01$ are suppressed for improved clarity.

\section{Error analysis}
The main sources of experimental errors are small uncontrolled fluctuations in the intensity of the transverse rf-field, non-idealities in its time modulation, and tiny inhomogeneities in the longitudinal static field as well as in the gradient pulses. All pulses in the experiment were optimized to minimize such errors. 

The error bars shown in the main text figures were Monte Carlo propagated, sampling deviations of the nuclear magnetization or quantum state tomography (QST) data with a Gaussian distribution having widths determined by the variances corresponding to such data. The variances of the magnetization measurements and QST data are obtained by preparing the same pseudo thermal state 100 times. These variances include random and systematic errors in both initial state preparation and data acquisition.

\newpage

\bibliography{references2}